\documentclass[12pt, twocolumn]{openjournal}

\usepackage{lipsum}

\usepackage{xcolor}
\usepackage{textgreek}
\usepackage[utf8]{inputenc}
\usepackage[english]{babel}
\usepackage{hyperref}
\hypersetup{
    colorlinks=true,
    linkcolor=blue,
    filecolor=blue,      
    urlcolor=red,
    citecolor=blue,
}
\usepackage{color,colortbl}
\usepackage{tensind}
\tensordelimiter{?}
\DeclareGraphicsExtensions{.bmp,.png,.jpg,.pdf}
\usepackage{verbatim}
\usepackage[normalem]{ulem}
\usepackage{orcidlink}
\usepackage{soul}
\usepackage{amsmath}
\usepackage{graphicx}

\begin{document}
\title{The stellar and dark matter distributions in early-type galaxies measured by stacked weak gravitational lensing}

\author{\vspace{-1.0cm}Momoka~Fujikawa$^{1}$}
\author{Masamune~Oguri$^{1, 2}$\orcidlink{0000-0003-3484-399X}}

\email{25wm2122@student.gs.chiba-u.jp}
\affiliation{$^{1}$Department of Physics, Graduate School of Science, Chiba University, 1-33 Yayoi-Cho, Inage-Ku, Chiba 263-8522, Japan}
\affiliation{$^{2}$Center for Frontier Science, Chiba University, 1-33 Yayoi-cho, Inage-ku, Chiba 263-8522, Japan}

\begin{abstract}
We investigate stellar mass and central dark matter density profiles of photometric luminous red galaxies with stellar masses of $\sim10^{10}-10^{12}M_\odot$ using weak gravitational lensing measurements from the Hyper Suprime-Cam Subaru Strategic Program data obtained with the Subaru Telescope. By stacking weak lensing signals from a large number of galaxies, we obtain average tangential shear profiles down to $\sim 10\,\mathrm{kpc}/h$, which are fitted assuming a two-component model consisting of stellar and dark matter components to constrain their central dark matter distribution. We find a preference for non-zero core radii of dark matter distributions in galaxies with stellar masses of $\sim 10^{11}M_\odot$. Our results imply a stronger feedback effect than that typically predicted by current hydrodynamical simulations. In addition, we provide a new constraint on the stellar-to-halo mass relation, where both stellar and halo masses are, for the first time, directly constrained by weak gravitational lensing. Our results prefer the stellar initial mass function (IMF) that is more bottom-heavy than the Salpeter IMF.
\end{abstract}


\maketitle

\section{Introduction}
\label{sec:intro}
The standard cosmological model is now established through both theoretical and observational approaches. In this standard model, the Cold Dark Matter (CDM) is assumed to be the form of dark matter. The CDM model neglects any interactions other than gravity in structure formation and assumes that the free streaming is ineffective. The free streaming refers to the suppression of density fluctuations caused by the free motion of dark matter particles with the large velocity dispersion. Since CDM does not experience such suppression, it allows the formation of structures down to very small scales.

While the standard cosmological model successfully explains large scale observations, it shows contradictions to some of smaller scale observations, such as those of dwarf galaxies. This is known as the small-scale challenges \citep{2017ARA&A..55..343B}.
One of these small-scale challenges is the core-cusp problem, which refers to the discrepancy between the central density profiles of dark matter halos predicted by CDM simulations and those measured from observations. The density distribution of a CDM halo is known to follow the Navarro-Frenk-White (NFW) profile \citep{1997ApJ...490..493N},
\begin{equation}
   \rho(r) = \frac{\rho_\mathrm{s}}{\frac{r}{r_\mathrm{s}}\left(1 + \frac{r}{r_\mathrm{s}}\right)^2},
\end{equation}
where $r$ is the distance from the halo center, $\rho$ is the mass density, $\rho_\mathrm{s}$ is the characteristic density, and $r_{\mathrm{s}}$ is the scale radius. In the inner region of the halo, where $r \ll r_\mathrm{s}$, the density behaves as $\rho(r) \propto r^{-1}$, indicating a cuspy profile. In contrast, several observational studies report cored dark matter distributions with nearly constant central densities \citep[e.g.,][]{2015AJ....149..180O}.

To address these small-scale challenges, various studies have been conducted at various mass scales. Among others, the central structure of early-type galaxies has been studied mainly using strong gravitational lensing and stellar dynamics, from which it has been shown that the total density profile follows an approximately isothermal form, $\rho(r)\propto r^{-2}$ \citep[e.g.,][]{2006ApJ...649..599K,2007ApJ...667..176G,2011MNRAS.415.2215B,2012ApJ...757...82B,2013MNRAS.432.1709C,2013ApJ...777...98S,2018MNRAS.480..431L,2023MNRAS.521.6005E}. Separating stellar and dark matter distributions requires breaking the degeneracy between the stellar initial mass function (IMF) and the dark matter fraction, which have been attempted using strong gravitational lensing \citep[e.g.,][]{2003ApJ...595...29R,2010ApJ...709.1195T,2010ApJ...721L.163A,2012ApJ...747L..15G,2014MNRAS.439.2494O,2019A&A...630A..71S,2025A&A...697A..95S}, stellar dynamics \citep[e.g.,][]{2012Natur.484..485C,2016ARA&A..54..597C,2017MNRAS.468.3949A,2022ApJ...930..153S,2023A&A...672A..84D,2024MNRAS.527..706Z,2024MNRAS.528.5295Y}, combining strong lensing and stellar dynamics
\citep[e.g.,][]{2004ApJ...611..739T,2015ApJ...800...94S,2015ApJ...814...26N,2018MNRAS.476..133O,2021MNRAS.503.2380S,2025MNRAS.541....1S}, and combining strong and weak lensing \citep[e.g.,][]{2007ApJ...667..176G,2018MNRAS.481..164S}. 

Recently, on the other hand, advances in observational techniques have made it possible to probe the inner density profiles of galaxies even through weak gravitational lensing. For instance, the possibility of probing the inner density profile of dwarf galaxies with stacked weak
lensing is considered in \citet{2015MNRAS.449.2128K}.
\citet{2024MNRAS.533..795K} used stacked weak lensing at
$>0.1~\mathrm{Mpc}$ to break the mass-sheet degeneracy inherent to the strong lensing analysis.

In this study, we conduct the first systematic analysis of the stellar mass and dark matter distributions in centers of early-type galaxies using stacked weak gravitational lensing. Since weak lensing probes the relatively outer regions of galaxy centers compared to methods using the velocity dispersion or strong lensing, our approach provides a means of studying central density profiles that is complementary to previous approaches. For this purpose, we employ the Hyper Suprime-Cam Subaru Strategic Program \citep[HSC-SSP;][]{2018PASJ...70S...4A} data that achieve both the wide area and the high number density of galaxies used for weak lensing.

This paper is organized as follows. In Section~\ref{sec:method}, we describe the HSC-SSP data used for our analysis as well as models to fit the data. We show our results in Section~\ref{sec:result}. We give our conclusion in Section~\ref{sec:concl}. Throughout the paper we assume a flat Universe with the matter density $\Omega_{\mathrm{m}}=0.3$, the cosmological constant $\Omega_{\Lambda}=0.7$, the baryon matter density $\Omega_{\mathrm{b}}=0.05$, the normalization of the matter power spectrum $\sigma_8=0.81$, the spectral index $n_{\mathrm{s}}=0.96$, and the dimensionless Hubble constant $h=0.7$.

\section{Data and Model}
\label{sec:method}

\subsection{Subaru HSC-SSP}\label{sec:hsc-ssp}

We use the HSC-SSP three-year shear catalog
\citep{2022PASJ...74..421L} for our weak lensing analysis. This
catalog covers an area of $\sim 430$~deg$^2$ with the raw galaxy
number density of $\sim 23$~arcmin$^{-2}$, and is validated with a
series of null tests for systematics. 

For a foreground lens sample, we employ the HSC-SSP final-year
photometric luminous red galaxy (LRG) sample \citep{2026PASJ..tmp....2O}.
The photometric LRG sample is constructed based on the stellar
population synthesis fitting of individual galaxies with the model of quiescent early-type galaxies based on \citet{2003MNRAS.344.1000B}, which is also used for the CAMIRA cluster finding algorithm
\citep{2014MNRAS.444..147O,2018PASJ...70S..20O,2018PASJ...70S..26O}.
The final-year photometric LRG sample covers an area of $\sim
1200$~deg$^2$ and contains galaxies in the photometric redshift range
of $0.05\leq z_{\mathrm{LRG}}\leq 1.25$ and the stellar mass range of
$M_\star\geq 10^{10.3}M_\odot$, where the stellar mass is derived assuming the
\citet{1955ApJ...121..161S} IMF. As shown in
\citet{2026PASJ..tmp....2O}, their photometric redshifts are accurate and precise
at $\sigma_z\lesssim 0.02$ in the redshift range of
$0.4 \lesssim z_{\mathrm{LRG}} \lesssim 1$.

We measure the differential surface density profile $\Delta\Sigma(r)$
around a sample of photometric LRGs with the standard approach
described in detail in e.g., \citet{2018PASJ...70S..25M} and
\citet{2018PASJ...70...30M}. Specifically, for each radial bin
centered at $r$, it is estimated as
\begin{equation}
  \Delta\Sigma(r)=\frac{1}{2\mathcal{R}}\frac{\sum_{\mathrm{l},\mathrm{s}}
    w_{\mathrm{ls}}
    \langle\Sigma_{\mathrm{cr}}^{-1}\rangle_{\mathrm{ls}}^{-1}(e_{+,\mathrm{ls}}-2\mathcal{R}c_{+,\mathrm{ls}})}{(1+K)\sum_{\mathrm{l},\mathrm{s}} w_{\mathrm{ls}}},
\end{equation}
where l and s denote the lens and the source, respectively,
$e_{+,\mathrm{ls}}$ is the tangential shear for each lens and source
pair, $c_{+,\mathrm{ls}}$ is the additive shear bias of the same
tangential shear component, the shear responsivity is
\begin{equation}
  \mathcal{R}=1-\frac{\sum_{\mathrm{l},\mathrm{s}}\sigma_{\mathrm{rms},\mathrm{s}}^2w_{\mathrm{ls}}}{\sum_{\mathrm{l},\mathrm{s}}w_{\mathrm{ls}}},
\end{equation}
with $e_{\mathrm{rms},\mathrm{s}}$ being the root-mean-square ellipticity
for the source s, $K$ is computed from the multiplicative shear bias
$m_{\mathrm{s}}$ as
\begin{equation}
  K=\frac{\sum_{\mathrm{l},\mathrm{s}}m_{\mathrm{s}}w_{\mathrm{ls}}}{\sum_{\mathrm{l},\mathrm{s}}w_{\mathrm{ls}}},
\end{equation}
and the weight factor $w_{\mathrm{ls}}$ is 
\begin{equation}
  w_{\mathrm{ls}}=\frac{\langle\Sigma_{\mathrm{cr}}^{-1}\rangle_{\mathrm{ls}}^2}{\sigma_{e,\mathrm{s}}^2+\sigma_{\mathrm{rms},\mathrm{s}}^2},
  \end{equation}
where $\sigma_{e,\mathrm{s}}$ is the measurement error of the
ellipticity and
$\langle\Sigma_{\mathrm{cr}}^{-1}\rangle_{\mathrm{ls}}$ is the average
inverse critical surface density computed from the probability
distribution function $P(z_{\mathrm{s}})$ of the photometric redshift of the source s as 
\begin{equation}
  \langle\Sigma_{\mathrm{cr}}^{-1}\rangle_{\mathrm{ls}}=\int_{z_{\mathrm{l}}}^\infty dz_{\mathrm{s}}\Sigma_{\mathrm{cr}}^{-1}(z_{\mathrm{l}},\,z_{\mathrm{s}})P(z_{\mathrm{s}}).
\end{equation}
We adopt DNNz photometric redshifts (A. J. Nishizawa et al., in prep.) for
photometric redshifts of source galaxies. The error on
$\Delta\Sigma(r)$ is derived from the weighted sum of ellipticity errors of
individual source galaxies, $\sigma_{e,\mathrm{s}}^2+\sigma_{\mathrm{rms},\mathrm{s}}^2$.
Since we are interested in lensing profiles at relatively small radii
where the intrinsic shape noise dominates
\citep[see e.g.,][]{2003MNRAS.339.1155H}, throughout the paper, we consider the
intrinsic shape noise only and ignore the cosmic shear noise. 

One of the most important sources of systematic errors, which is
relevant particularly at small radii, is the dilution effect by
neighboring galaxies. Due to clustering of galaxies, there is a higher
probability of having their redshifts almost same as that of a lensing
object, which is usually not accounted for in deriving $P(z_{\mathrm{s}})$
and hence biases lensing measurements. As shown in e.g.,
\citet{2018PASJ...70...30M}, this dilution effect can be mitigated by
carefully selecting background source galaxies and using only them for
the weak lensing analysis. In this paper, we employ the so-called
$P(z)$ cut method \citep{2014MNRAS.444..147O} for which we only use
source galaxies that satisfy 
\begin{equation}
\int_{z_{\mathrm{l}}+\Delta z}^\infty dz_{\mathrm{s}}\,P(z_{\mathrm{s}})>P_{\mathrm{cut}}.
\label{eq:Pcut}
\end{equation}
While we fix $\Delta z=0.1$ and $P_{\mathrm{cut}}=0.95$, in
Appendix~\ref{ap:ap} we explicitly check the robustness of our result
with respect to the choice of the value of $P_{\mathrm{cut}}$.

\subsection{Model of the inner density profile}

For fitting the inner region of the observed differential surface density distribution, we use a two-component model consisting of a stellar matter component and a dark matter component.
For the stellar matter component, we adopt the Hernquist density profile \citep{1990ApJ...356..359H}
\begin{equation}
  \rho(r) = \frac{M_\star}{2\pi}\frac{a}{r}\frac{1}{\left(r+a\right)^3},
  \label{eq:rhosm}
\end{equation}
where $M_\star$ is the stellar mass. The scale radius $a$ is related to the effective (half-light) radius $r_\mathrm{e}$ as
\begin{equation}
    r_\mathrm{e} \approx 1.8153a.
\end{equation}
The mass enclosed within a radius $r$ is given by the following equation
\begin{equation}
  M_{\star,\mathrm{3D}}(<r) = M_\star\frac{r^2}{\left(r+a\right)^2}.
  \label{eq:m2dsm}
\end{equation}
Analytic expressions of lensing profiles of the Hernquist profile are given in \citet{2001astro.ph..2341K}.

For the dark matter component, we consider a model that has a core at the center and follows a power law in the outer region. We define the three-dimensional density profile as
\begin{equation}
  \rho_{\mathrm{DM}}\left(r\right)
  = M_\star\frac{3+\gamma}{2\pi^{3/2} \, r_\mathrm{e}^{3}} \frac{\Gamma\!\left(-\tfrac{\gamma}{2}\right)}{\Gamma\!\left(-\tfrac{1+\gamma}{2}\right)}A\left( \frac{r^{2} + r_\mathrm{c}^{2}}{r_\mathrm{e}^{2}} \right)^{\gamma/2},
\end{equation}
where $\Gamma(x)$ is the gamma function, $\gamma$ is the radial slope of the density profile, $A$ is the dimensionless parameter that determines the normalization, and $r_\mathrm{c}$ is the core radius. From the definition of the surface mass density, we obtain
\begin{align}
  \Sigma_{\mathrm{DM}}\left(r\right) =&\int_{-\infty}^{\infty}\rho_{\mathrm{DM}}\!\left(\sqrt{r^{2}+Z^{2}}\right)dZ\\
  =& M_\star\frac{3+\gamma}{2\pi r_\mathrm{e}^{3+\gamma}}A\left( r^{2} + r_\mathrm{c}^{2}\right)^{\frac{1+\gamma}{2}}.
\end{align}
Furthermore, the two-dimensional enclosed mass can be expressed in terms of the surface mass density as
\begin{align}
  M_{\mathrm{DM},2\mathrm{D}}\left(<r\right) =&\int_{0}^{r} 2\pi r'\Sigma_{\mathrm{DM}}\left(r'\right)dr'\\
  =& M_\star A\left(\frac{r_\mathrm{c}}{r_\mathrm{e}}\right)^{3+\gamma}
  \left[\left(1+\frac{r^2}{r_\mathrm{c}^2}\right)^{\tfrac{3+\gamma}{2}}-1\right].
\end{align}
In addition, the average surface mass density within the radius $r$ is given by
\begin{align}
  \bar{\Sigma}_{\mathrm{DM}} \left(<r\right)=&\frac{M_{\mathrm{DM},2\mathrm{D}}\left(<r\right)}{\pi r^{2}}\\
  =& \frac{M_\star A}{\pi r^2}\left(\frac{r_\mathrm{c}}{r_\mathrm{e}}\right)^{3+\gamma}
  \left[\left(1+\frac{r^2}{r_\mathrm{c}^2}\right)^{\tfrac{3+\gamma}{2}}-1\right].
\end{align}
From these results, the differential surface mass density is computed as
\begin{align}
  \Delta \Sigma_{\mathrm{DM}}\left(r\right) =&\bar{\Sigma}_{\mathrm{DM}}\left(<r\right) - \Sigma_{\mathrm{DM}}\left(r\right)\\
  =&\frac{M_\star A}{\pi r_\mathrm{e}^{3+\gamma}}\left\{\frac{1}{r^{2}}\left[\left(r^{2}+r_\mathrm{c}^{2}\right)^{\frac{3+\gamma}{2}}-r_\mathrm{c}^{3+\gamma}\right]\right.\notag\\
  &-\left.\frac{3+\gamma}{2}\left(r^{2}+r_\mathrm{c}^{2}\right)^{\frac{1+\gamma}{2}}\right\}.
\end{align}
We note that this cored power-law model for the dark matter distribution is rather empirical and is not based on any specific dark matter or baryon feedback models. We choose the cored power-law model based on that the fact the lensing profiles can be calculated analytically and also that it is a sufficiently flexible model of the inner dark matter density profile. We note that our analysis of the inner profile fitting is restricted within a relatively narrow fitting range as described in Section~\ref{subsec:method}, within which the cored power-law model can approximate a wide variety of different physically-motivated models. For comparison, previous strong lensing studies that adopted a two-component model assume e.g., a power-law profile, a generalized NFW profile without a core, or a cored power-law profile \citep[e.g.,][]{2003ApJ...595...29R,2014MNRAS.439.2494O,2020OJAp....3E..10W,2025MNRAS.541....1S,2025A&A...697A..95S}. Our cored power-law model is comparably or more flexible than those adopted in previous strong lensing studies when compared within the narrow fitting range in radius. Since the cored power-law model is intended for the model of the inner density profile only, we caution that it does not provide a good fit to the observed dark matter distribution when it is extrapolated to the outer part of the density profile.

We fit the sum of the two components 
\begin{equation}
    \Delta \Sigma\left(r\right) = \Delta\Sigma_{\star}\left(r\right)+\Delta \Sigma_{\mathrm{DM}}\left(r\right),
    \label{eq:deltasigma}
\end{equation}
to the observational data.

\subsection{Model of the outer density profile}
\label{subsec:outer model}

While our main focus is the inner density profile, we also fit the
outer lensing profile to estimate the halo mass and compare our
analysis result of the inner density profile with an extrapolation 
of the NFW profile fitted to the outer
lensing profile. Here a complication lies in the fact that all the
photometric LRGs are not necessarily central galaxies within halos, but
can correspond to satellite galaxies in the outer part of massive
halos. We thus fit the outer lensing profile to the following simple model
\begin{align}
  \Delta \Sigma(r) =& \Delta\Sigma_{\mathrm{cen}}(r;\,M)+
  f_{\mathrm{sat}}\Delta\Sigma_{\mathrm{sat}}(r;\,M_{\mathrm{h}})\nonumber\\
  &+\Delta\Sigma_{\mathrm{2h}}(r;\,M,\,M_{\mathrm{h}}),
  \label{eq:NFW}
\end{align}
for which we assume that photometric LRGs are hosted by halos with the
average mass $M$ and the fraction $f_{\mathrm{sat}}$ of photometric
LRGs corresponds to satellite galaxies with the average host mass
$M_{\mathrm{h}}$. 
We do not multiply $(1-f_{\mathrm{sat}})$ in the $\Delta\Sigma_{\mathrm{cen}}$ term assuming that subhalos hosting satellite galaxies also contribute to $\Delta\Sigma_{\mathrm{cen}}$, consistently with weak lensing observations of satellite galaxies  \citep[e.g.,][]{2022MNRAS.517.4389K}.
Following the analysis of ray-tracing simulations
presented in \citet{2011MNRAS.414.1851O}, we model
$\Delta\Sigma_{\mathrm{cen}}(r;\,M)$ by a smoothly truncated NFW density
profile \citep{2009JCAP...01..015B} with the truncation radius 
$r_{\mathrm{t}}/r_{\mathrm{vir}}=2.5$. We model the satellite
component as
\begin{equation}
  \Delta\Sigma_{\mathrm{sat}}(r;\,M_{\mathrm{h}})
  =\int \frac{k\,dk}{2\pi}
  \frac{\left\{\tilde{\Sigma}(k;\,M_{\mathrm{h}})\right\}^2}{M_{\mathrm{h}}}J_2(kr),
\end{equation}
where $\tilde{\Sigma}$ is the Fourier transform of the
\citet{2009JCAP...01..015B} density profile and $J_2(x)$ is the Bessel
function of order 2. 
This expression of $\Delta\Sigma_{\mathrm{sat}}$ properly takes account of the off-centering effect on the lensing profile assuming that the number density profile of satellite galaxies follows the density profile of the host halo \citep[see e.g.,][]{2013MNRAS.435.2345H}.
For both $\Delta\Sigma_{\mathrm{cen}}(r;\,M)$ and
$\Delta\Sigma_{\mathrm{sat}}(r;\,M_{\mathrm{h}})$, we adopt the
mass-concentration relation of \citet{2015ApJ...799..108D} and
\citet{2019ApJ...871..168D} implemented in the {\tt colossus} package
\citep{2018ApJS..239...35D}. We compute the 2-halo term as
\begin{equation}
  \Delta\Sigma_{\mathrm{2h}}(r;\,M,\,M_{\mathrm{h}})
  =\int \frac{k\,dk}{2\pi}\bar{\rho}_{\mathrm{m0}}\bar{b}P_{\mathrm{m}}(k/(1+z);\,z)J_2(kr),
\end{equation}
where $\bar{\rho}_{\mathrm{m0}}$ is the average matter density of the
Universe at $z=0$, $P_{\mathrm{m}}(k/(1+z);\,z)$ is the linear
matter power spectrum, and the average halo bias $\bar{b}$ is
\begin{equation}
  \bar{b}=(1-f_{\mathrm{sat}})b(M)+f_{\mathrm{sat}}b(M_{\mathrm{h}}),
\end{equation}
with $b(M)$ being computed by the model of \citet{2010ApJ...724..878T}.

Throughout the paper, we ignore any contribution from the hot gas. For the inner profile fitting, X-ray observations suggest that the contribution of the hot gas to the total mass profile is $\lesssim 1$\% \citep[e.g.,][]{2003ARA&A..41..191M,2006ApJ...636..698F} and hence can be safely neglected. Since the gas mass fraction increases with increasing radius, the halo mass derived from the outer profile fitting may be interpreted as the total mass including the gas mass.

\begingroup 
    \setlength{\tabcolsep}{10pt} 
    \renewcommand{\arraystretch}{1.5} 
    \begin{table*}[t]
        \centering
        \begin{tabular}{ cccccccc }  
    \hline \hline
$\log_{10}(M_{\star,\mathrm{in}}[M_\odot])$ & $N_{\mathrm{LRG}}$ & $r_{\mathrm{e}}$ [$\mathrm{Mpc}/h$]& 
$A$ & $\gamma$ & $r_{\mathrm{c}}$ [$\mathrm{Mpc}/h$] & $M_{\star,\mathrm{fit}}$ [$M_\odot / h$]
& $\chi^2_\mathrm{in}/\mathrm{dof}$ \\
            \hline     
10.3--10.5 & $1.81\times 10^5$ & $1.4 \times 10^{-3}$ & $7.5^{+12.0}_{-6.0}$ & $-2.50^{+0.50}_{-0.00}$ & $1.8^{+0.4}_{-0.9} \times 10^{-2}$ & $3.0^{+1.5}_{-2.2} \times 10^{10}$ & 9.4/5 \\
            \hline
10.5--10.7 & $2.05\times 10^5$ & $1.4 \times 10^{-3}$ & $0.2^{+4.0}_{-0.0}$ & $-1.51^{+0.06}_{-0.29}$ & $0.0^{+1.5}_{-0.0} \times 10^{-2}$ & $2.5^{+4.1}_{-2.3} \times 10^{10}$ & 10.2/5 \\
            \hline
10.7--10.9 & $2.23\times 10^5$ & $1.4 \times 10^{-3}$ & $2.7^{+0.4}_{-2.1}$ & $-2.50^{+0.33}_{-0.00}$ & $1.7^{+0.3}_{-0.4} \times 10^{-2}$ & $9.0^{+1.5}_{-1.5} \times 10^{10}$ & 11.8/5 \\
            \hline
10.9--11.1 & $2.08\times 10^5$ & $1.5 \times 10^{-3}$ & $1.9^{+0.3}_{-1.6}$ & $-2.50^{+0.41}_{-0.00}$ & $2.4^{+0.5}_{-0.6} \times 10^{-2}$ & $1.7^{+0.2}_{-0.1} \times 10^{11}$ & 8.8/5 \\
            \hline
11.1--11.3 & $1.49\times 10^5$ & $1.8 \times 10^{-3}$ & $1.9^{+0.7}_{-1.8}$ & $-2.46^{+0.76}_{-0.04}$ & $2.5^{+0.5}_{-1.3} \times 10^{-2}$ & $2.2^{+0.1}_{-0.2} \times 10^{11}$ & 19.2/5 \\
            \hline
11.3--11.5 & $5.54\times 10^4$ & $2.3 \times 10^{-3}$ & $0.2^{+3.9}_{-0.1}$ & $-1.83^{+0.52}_{-0.67}$ & $1.8^{+1.7}_{-1.8} \times 10^{-2}$ & $2.7^{+0.4}_{-1.3} \times 10^{11}$ & 11.1/5 \\
            \hline
11.5--11.7 & $1.02\times 10^4$ & $3.3 \times 10^{-3}$ & $0.2^{+8.0}_{-0.1}$ & $-1.56^{+0.55}_{-0.52}$ & $1^{+16}_{-1} \times 10^{-3}$ & $2.7^{+1.6}_{-2.5} \times 10^{11}$ & 6.4/5 \\
            \hline \hline  
        \end{tabular}  
        \caption{Parameters used in the inner profile fitting. $N_{\mathrm{LRG}}$ denotes the number of LRGs used as lens galaxies, and $r_{\mathrm{e}}$ represents the half-light radius. The errors are shown at the $1\sigma$ level.}
        \label{table:in}
    \end{table*}
\endgroup

\subsection{Fitting method}
\label{subsec:method}

For the lensing profile fittings, we perform a chi-square analysis using $\chi^2$ defined by
\begin{align}
  \chi^2 = \sum_{i=1}^{N} \frac{\left[\Delta\Sigma_{\mathrm{obs},i} - f_{\mathrm{model}}(r_i)\right]^2}{\sigma_{i}^2},&\\
  f_{\mathrm{model}}(r)=\frac{\Delta\Sigma(r)}{1-{\langle\Sigma_{\mathrm{cr}}}^{-1}\rangle\Sigma(r)}&.
  \label{eq:chi2}
\end{align}
Here, $\Delta\Sigma_{\mathrm{obs}}$ and $\sigma$ are the differential surface mass density and its error obtained from the weak lensing measurement, respectively, $i$ labels the radial bin, and $\Delta\Sigma$ and $\Sigma$ are the model predictions based on the inner or outer density profile mentioned above. We ignore off-diagonal terms of the covariance matrix, which is justified by the fact that off-diagonal terms are negligibly small within the range of the radial bins of our interest where the shape noise is dominated \citep[e.g.,][see also discussion in Section~\ref{sec:hsc-ssp}]{2017MNRAS.471.3827S,2019PASJ...71..107M,2019ApJ...875...63M}. In computing $f_{\mathrm{model}}(r)$, we include the correction of the surface mass density profile to take account of the fact that observed lensing shear profiles are reduced shear. The errors on the parameters are estimated from the distribution of
\begin{equation}
  \Delta\chi^2(x) = \chi^2_\mathrm{min}(x)-\chi^2_\mathrm{min},
  \label{eq:error}
\end{equation}
where $\chi^2_\mathrm{min}(x)$ is the minimum $\chi^2$ value when a parameter $x$ is fixed, and $\chi^2_\mathrm{min}$ is the minimum $\chi^2$ value when all parameters are allowed to vary. When computing $\Delta\chi^2$, we compute all $\chi^2$ values in grid points of the full parameter space to search for the minimum $\chi^2$ and obtain the marginalized constraint on each parameter.

We divide the stellar mass of LRGs into seven bins in the range $10^{10.3}M_\odot<M_{\star}<10^{11.7}M_\odot$, and measure the lensing profile for each bin. We restrict our analysis to the lens redshift range of $0.4<z<0.6$. This choice of the redshift range is motivated by several reasons, including the fact that the lensing signal is strongest in this redshift range given the depth of the HSC-SSP, the photometric redshift accuracy of photometric LRGs is best, and at lower redshifts, high surface brightnesses of lensing galaxies can obscure background galaxies or bias their shape measurements near centers of lensing galaxies.

For the inner lensing profile fitting, we treat $M_{\star,\mathrm{fit}}$, $A$, $\gamma$, and $r_\mathrm{c}$ as parameters. Here, the stellar mass $M_{\star,\mathrm{fit}}$ is treated as a free parameter independently of observed stellar masses of photometric LRGs, which we denote $M_{\star,\mathrm{in}}$ hereafter, because the true stellar mass can significantly differ from $M_{\star,\mathrm{in}}$ considering various effects such as the uncertainty of the stellar IMF, systematic errors in inferring stellar masses with the stellar population synthesis fitting, and systematic errors in measuring total magnitudes of bright galaxies in the HSC-SSP data \citep{2018MNRAS.475.3348H}. We calculate the effective radius $r_{\mathrm{e}}$ for each $M_{\star,\mathrm{in}}$ bin based on \citet{2025OJAp....8E...8A}, which adopts the model of \citet{2024ApJ...960...53V}, and ignore its uncertainty because our fitting range of radii (see below) is typically much larger than $r_{\mathrm{e}}$ and hence its uncertainty does not significantly affect our result. The values of $r_{\mathrm{e}}$ are summarized in Table~\ref{table:in}.  We exclude the very central region from the fitting range because the weak lensing signal in that range suffers from large statistical and systematic errors (see Appendix~\ref{ap:P}). The fitting is therefore performed within the range $0.013\,\mathrm{Mpc}/h<r<0.1\,\mathrm{Mpc}/h$. While we find that the off-centered profile $\Delta\Sigma_{\mathrm{sat}}$ and the two-halo term $\Delta\Sigma_{\mathrm{2h}}$ in Equation~\eqref{eq:NFW} contribute little in this fitting range, for definiteness we subtract these contributions based on the best-fitting model of the outer profile fitting (see below) from the observed differential surface mass density profile when we perform the inner profile fitting. In addition, since values of the core radius $r_\mathrm{c}$ larger than this fitting range are not physically meaningful, we set its upper limit to $0.1\,\mathrm{Mpc}/h$. 
We note that, even though we fit lensing profiles down to quite small scales, observed reduced shear values in this fitting range are $g\lesssim 0.1$, which is well within the range where current shape measurement techniques can accurately measure weak lensing shear \citep[e.g.,][]{2016MNRAS.461.3794O,2020A&A...640A.117H} and roughly within the range tested with cluster weak lensing using the HSC-SSP shape catalog \citep[e.g.,][]{2018PASJ...70...30M}.

For the outer lensing profile fitting, we treat $M$, $M_\mathrm{h}$, and $f_\mathrm{sat}$ as parameters. The fitting range is set from $r=0.04\,\mathrm{Mpc}/h$, where the $\Delta\Sigma_\mathrm{cen}$ component begins to dominate, to $r=3\,\mathrm{Mpc}/h$, where the effect of cosmic shear on the statistical error of the tangential shear measurement can be considered negligible. Also, we set the upper limit of $f_\mathrm{sat}$ to 0.4 given the constraint on $f_\mathrm{sat}$ from the halo-model analysis of the clustering of photometric LRGs \citep{2021ApJ...922...23I}.

\begin{figure*}[tbp]
  \centering
    \includegraphics[width=0.245\linewidth]{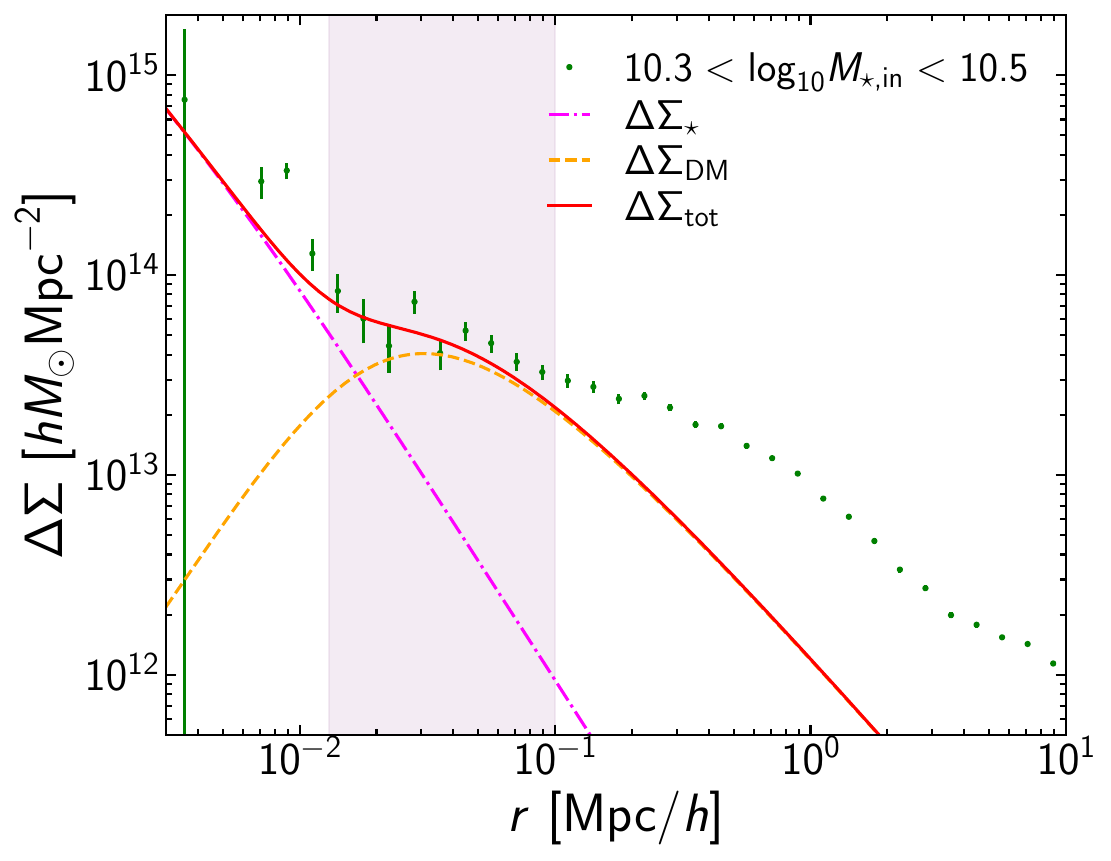}
    \includegraphics[width=0.245\linewidth]{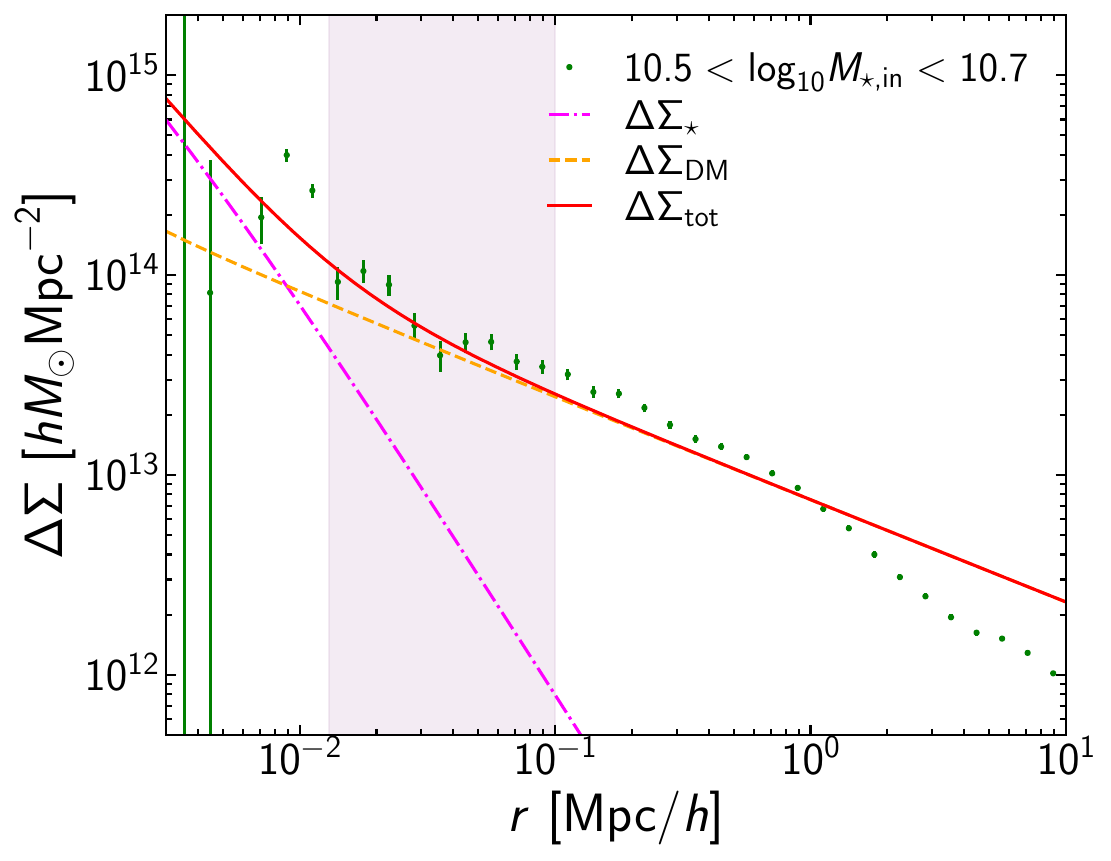}
    \includegraphics[width=0.245\linewidth]{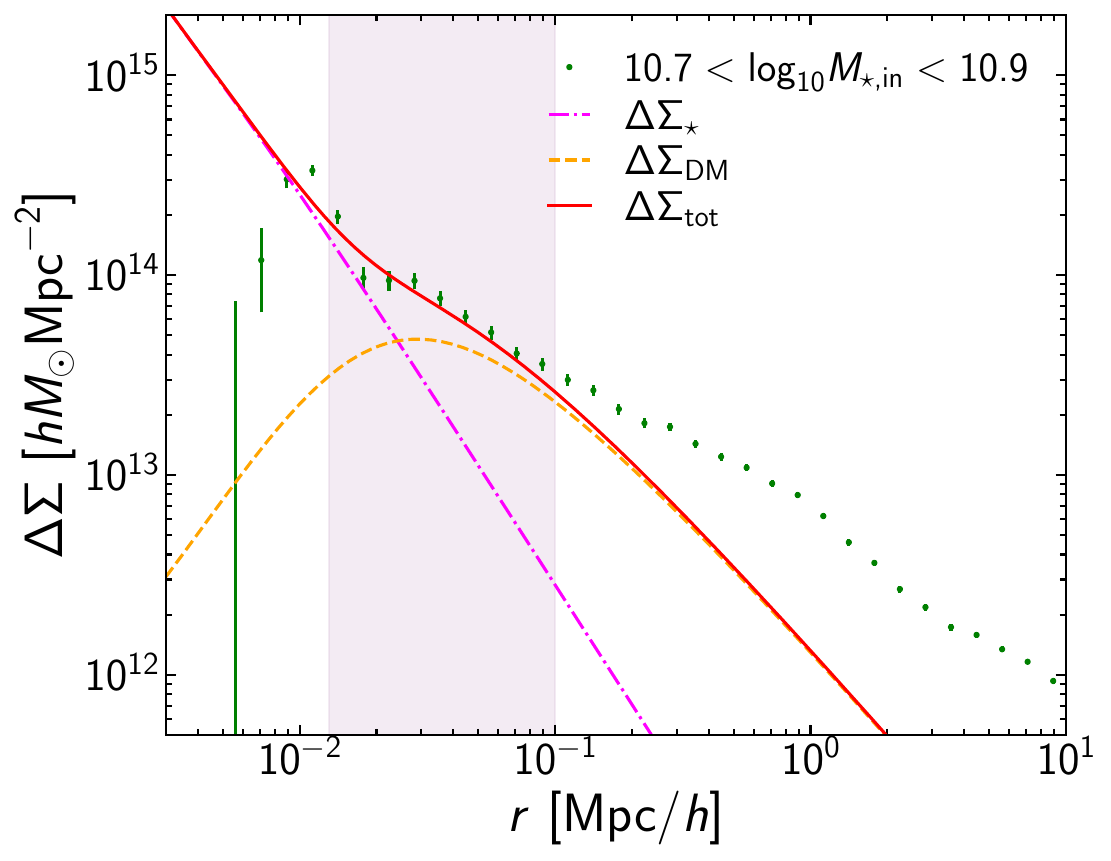}
    \includegraphics[width=0.245\linewidth]{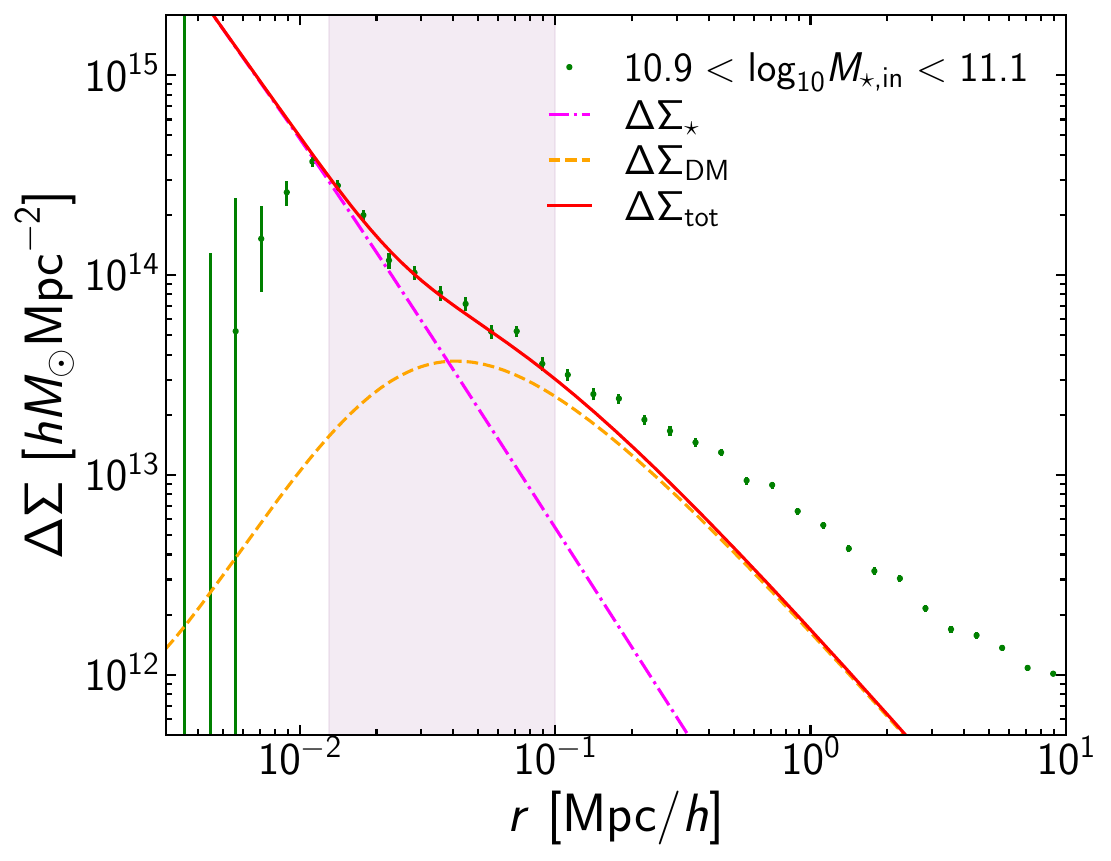}

  \vskip\baselineskip

  \begin{flushleft}
      \includegraphics[width=0.245\linewidth]{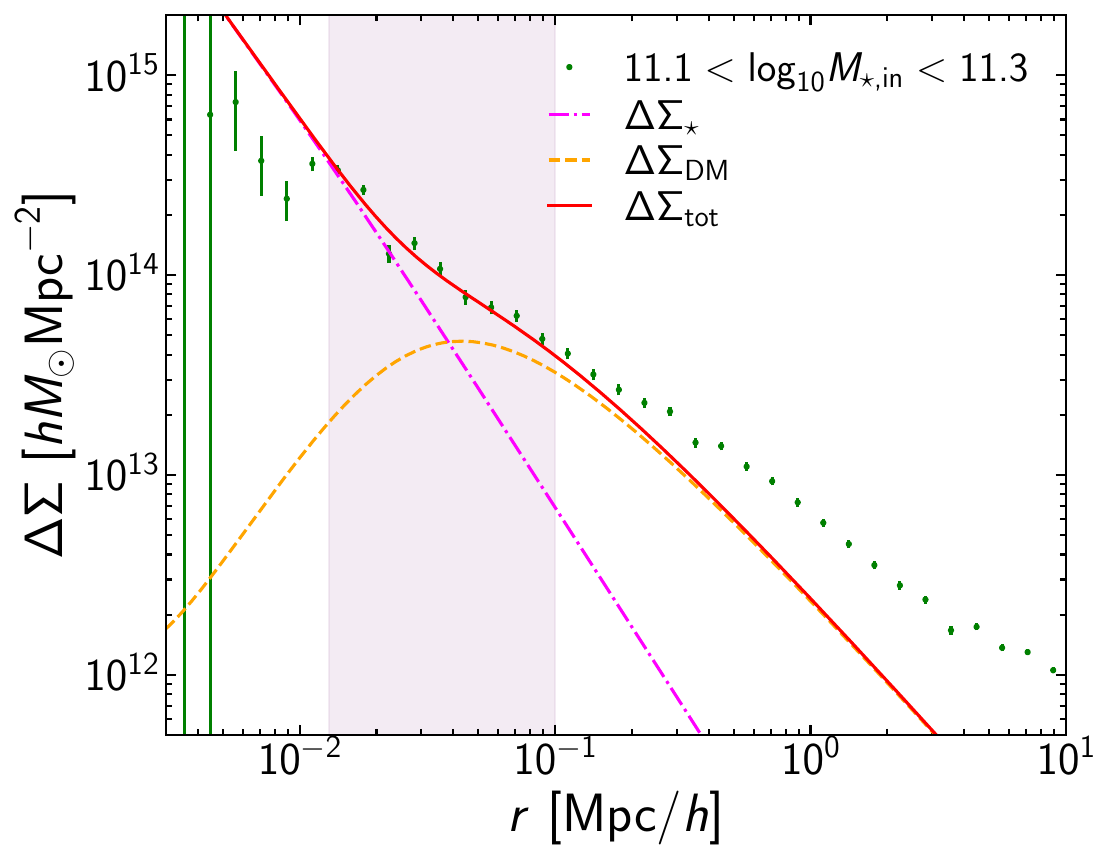}
      \includegraphics[width=0.245\linewidth]{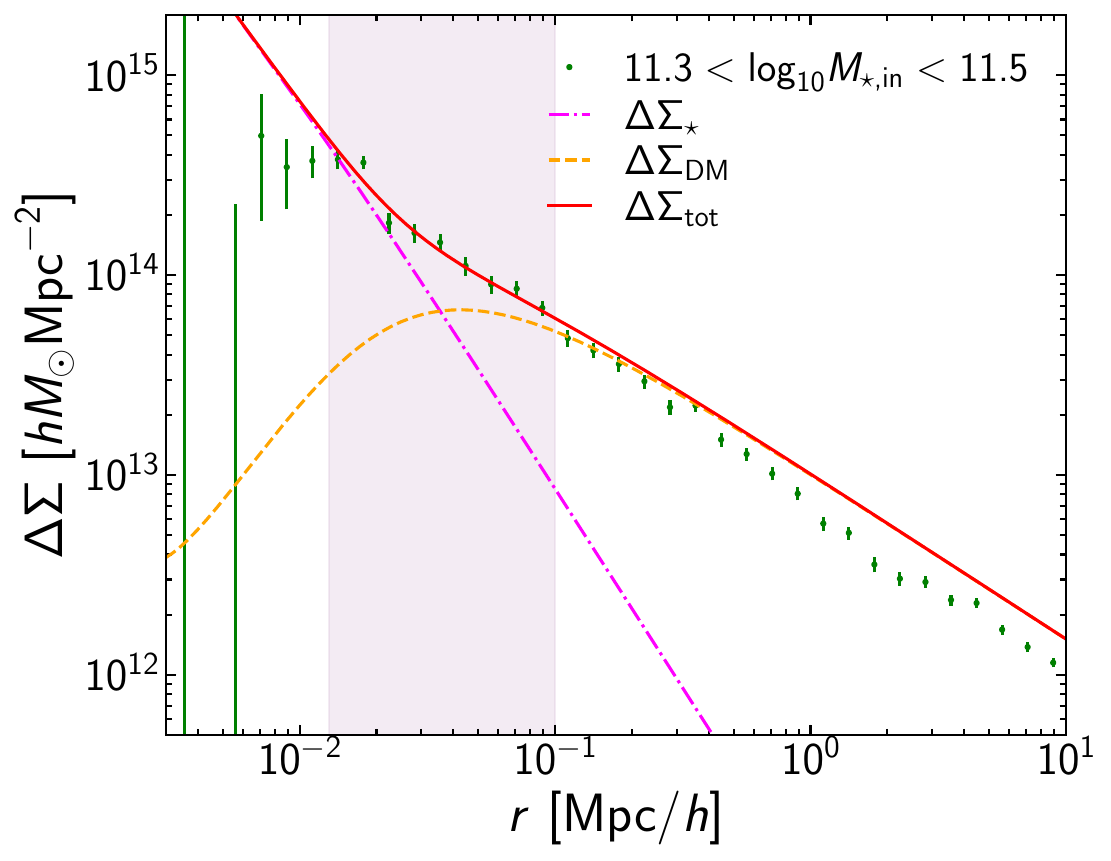}
      \includegraphics[width=0.245\linewidth]{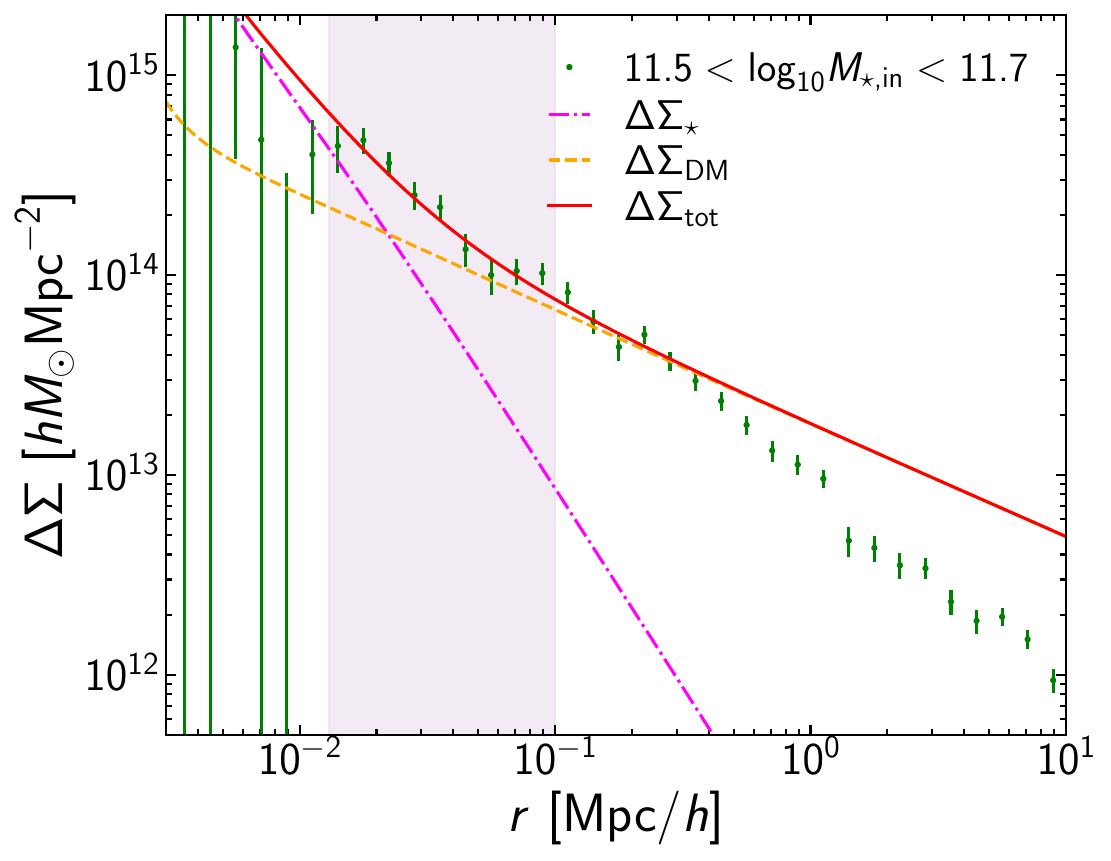}
  \end{flushleft}

  \caption{Surface mass density profiles $\Delta\Sigma$ obtained from weak lensing measurements and the results of fitting the inner profiles. Different panels show results for different stellar mass bins. The green points show the observational results from stacked weak gravitational lensing. The magenta dash-dotted, the orange dashed, and the red solid lines indicate the stellar matter component, the dark matter component, and the total mass, respectively. The shaded region denotes the fitting range.
}
  \label{fig:inner fit}
\end{figure*}

\begin{figure}[tbp]
  \includegraphics[width=\linewidth]{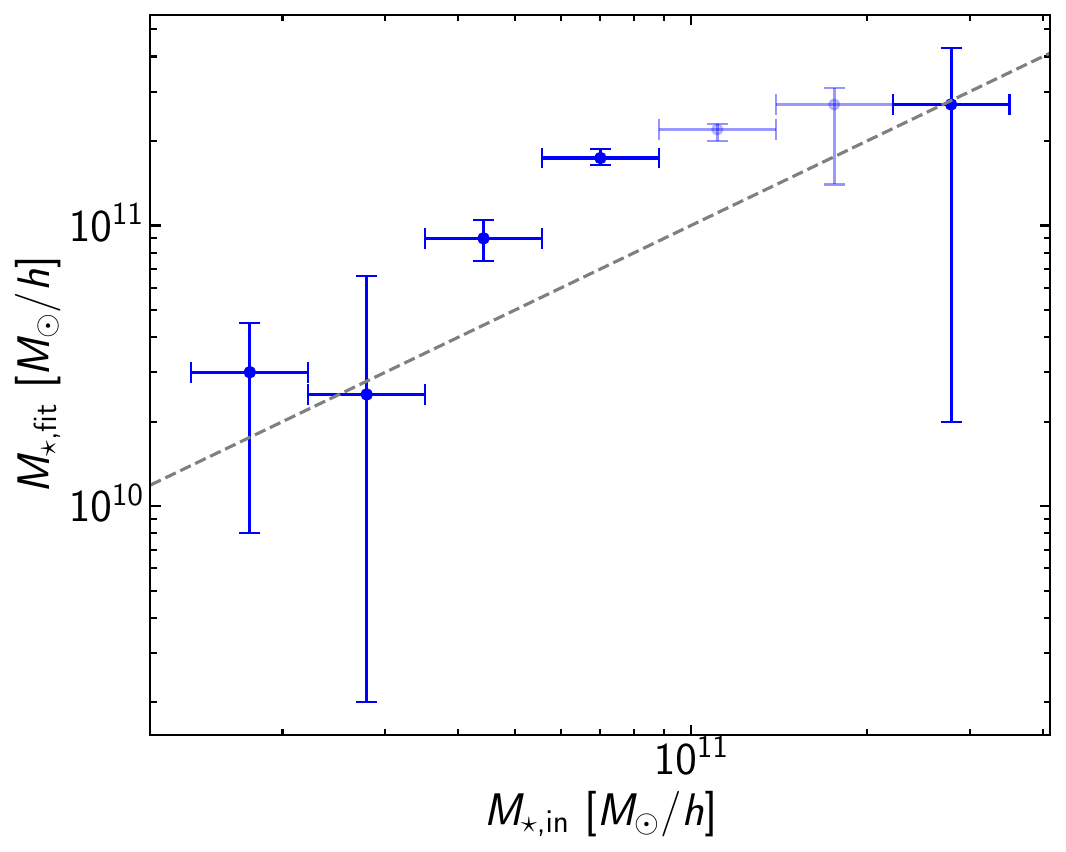}
  \caption{Comparison of best-fit stellar masses $M_{\star,\mathrm{fit}}$ with those measured from the HSC-SSP photometric data, $M_{\star,\mathrm{in}}$. The blue points show our results and the dashed line shows $M_{\star,\mathrm{fit}}=M_{\star,\mathrm{in}}$. Points shown by the lighter color might be affected by systematic errors and therefore should be interpreted with caution (see the text for more details).}
  \label{fig:min}
\end{figure}

\begin{figure*}[tbp]
  \centering
    \includegraphics[width=0.245\linewidth]{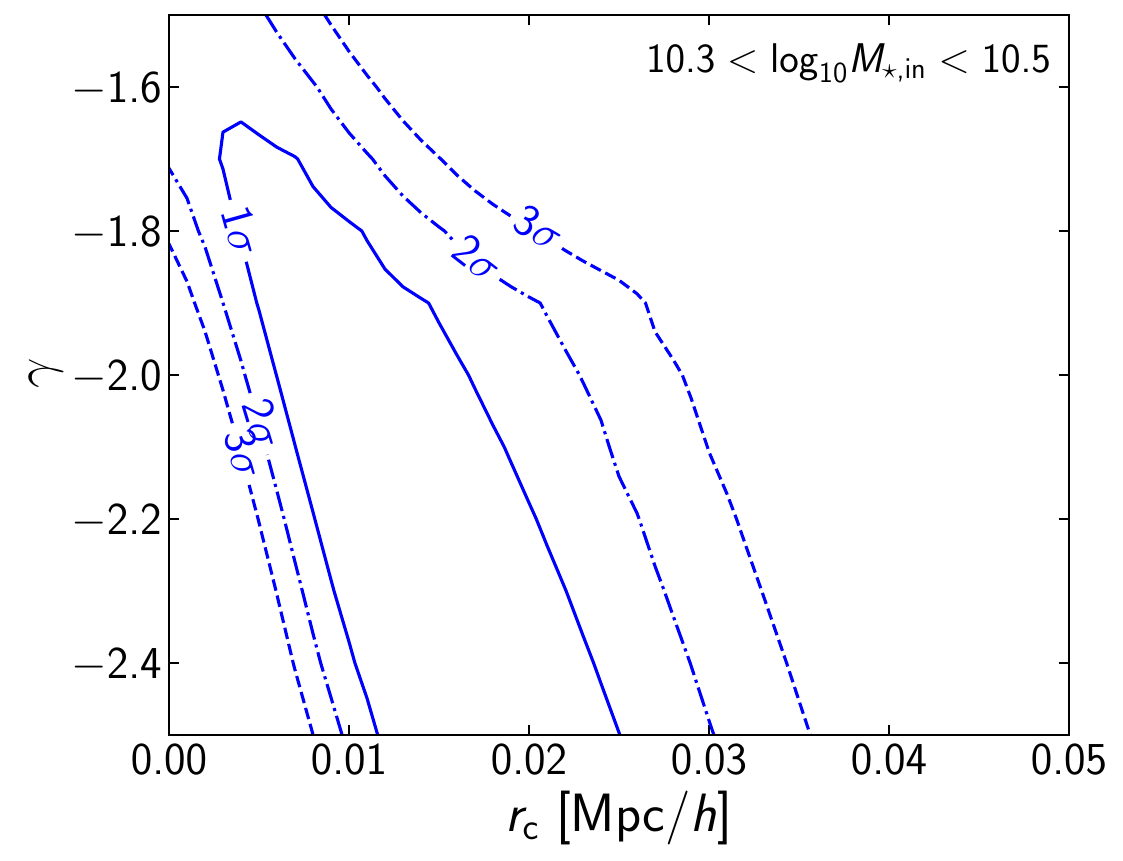}
    \includegraphics[width=0.245\linewidth]{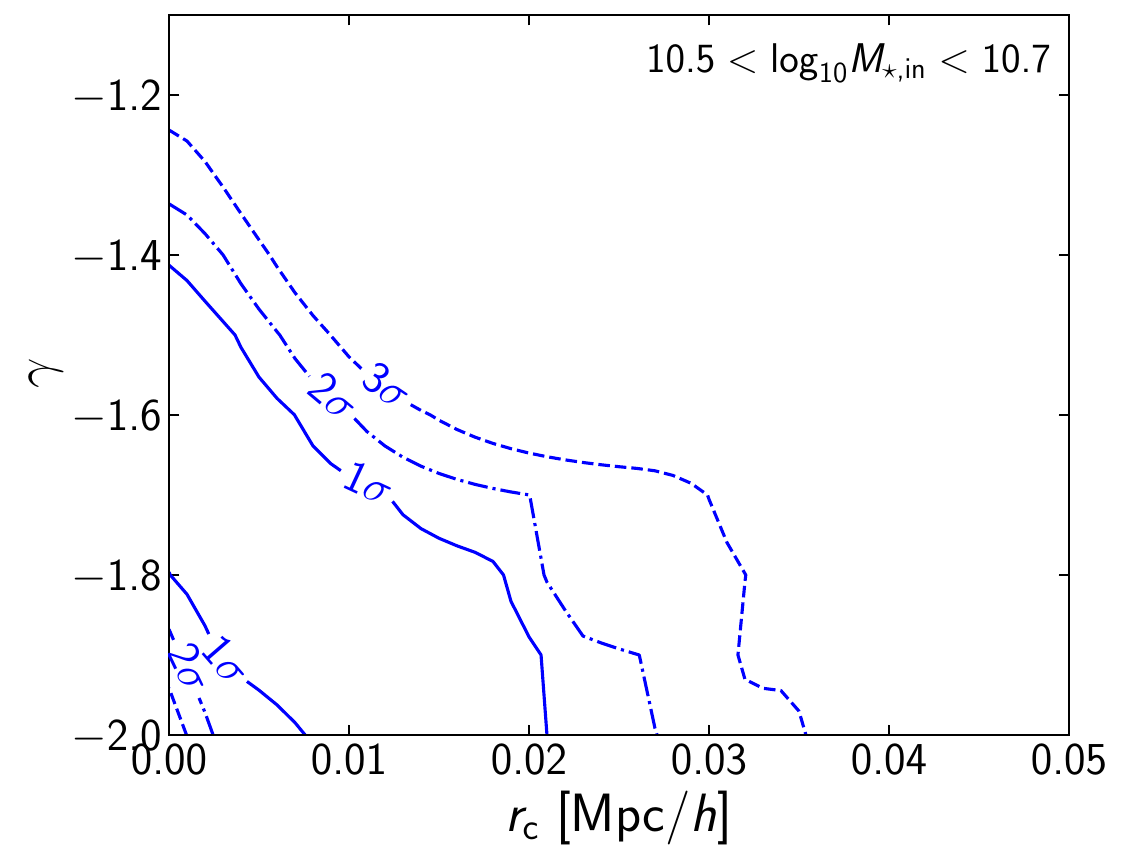}
    \includegraphics[width=0.245\linewidth]{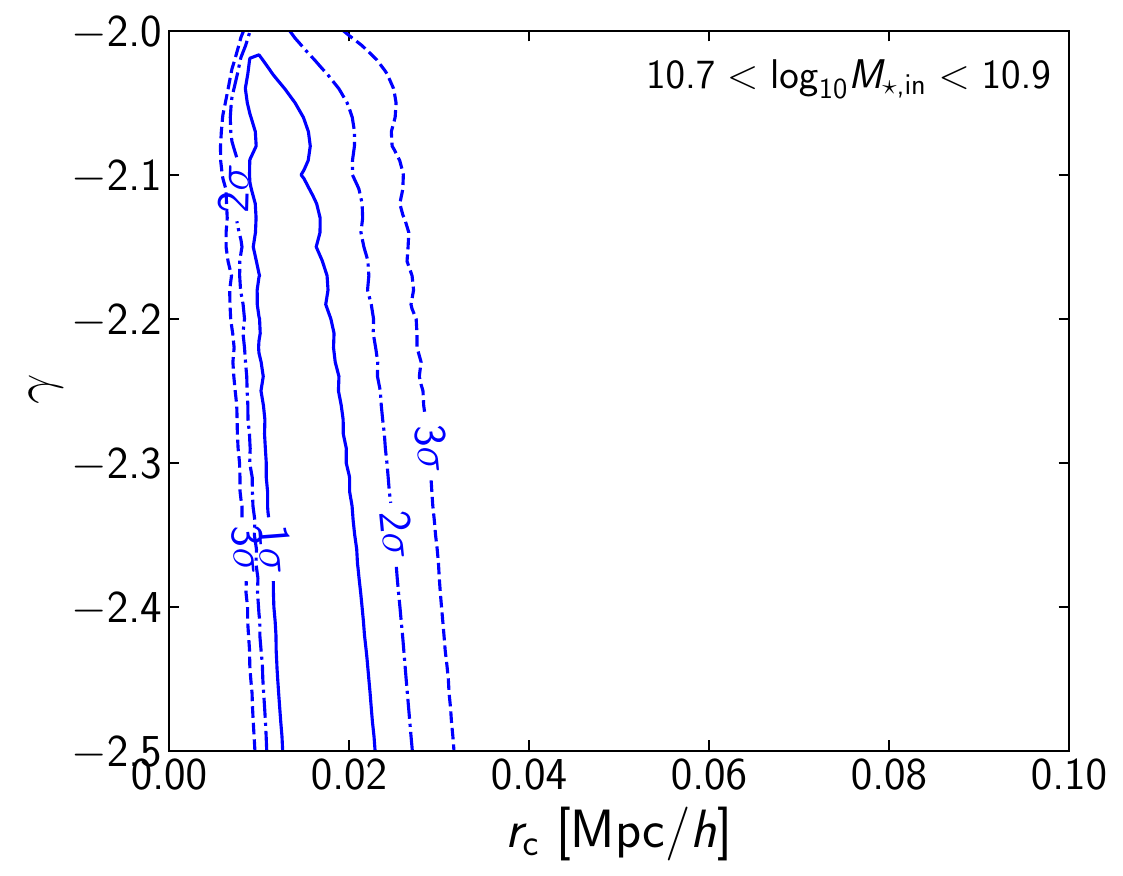}
    \includegraphics[width=0.245\linewidth]{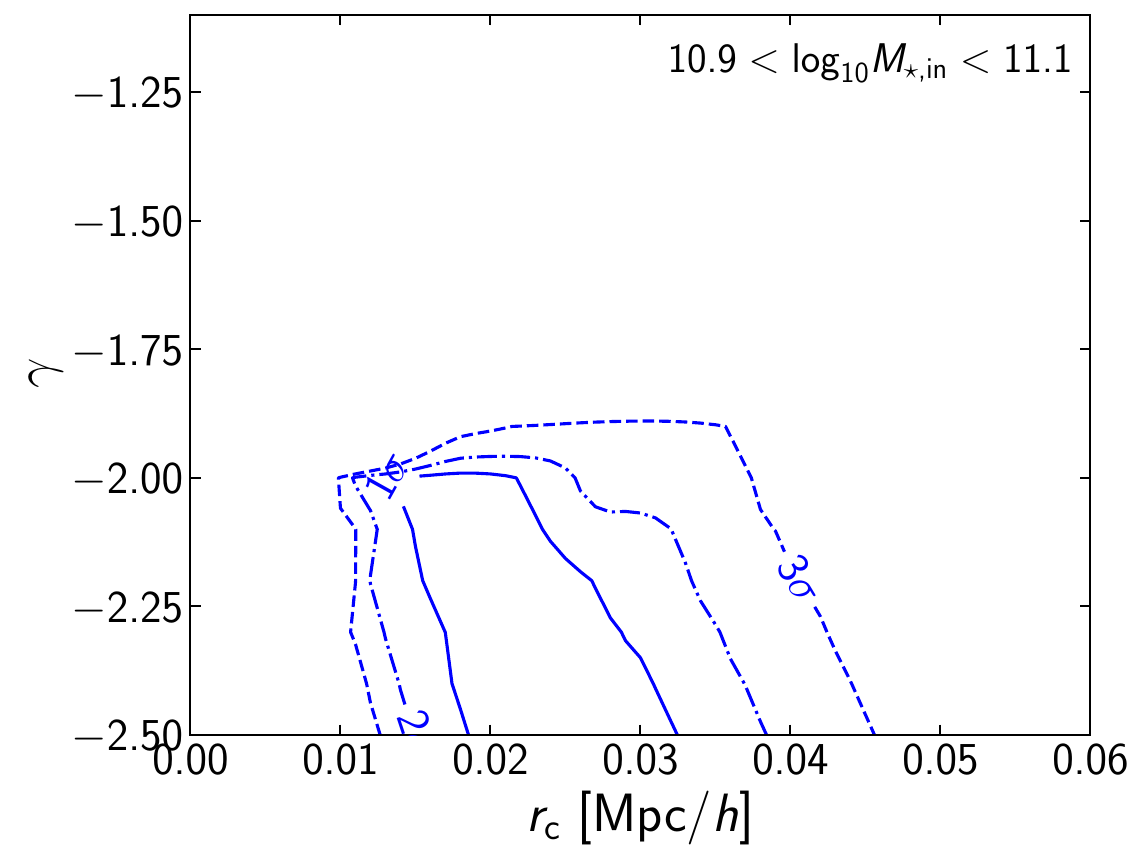}

  \vskip\baselineskip

  \begin{flushleft}
      \includegraphics[width=0.245\linewidth]{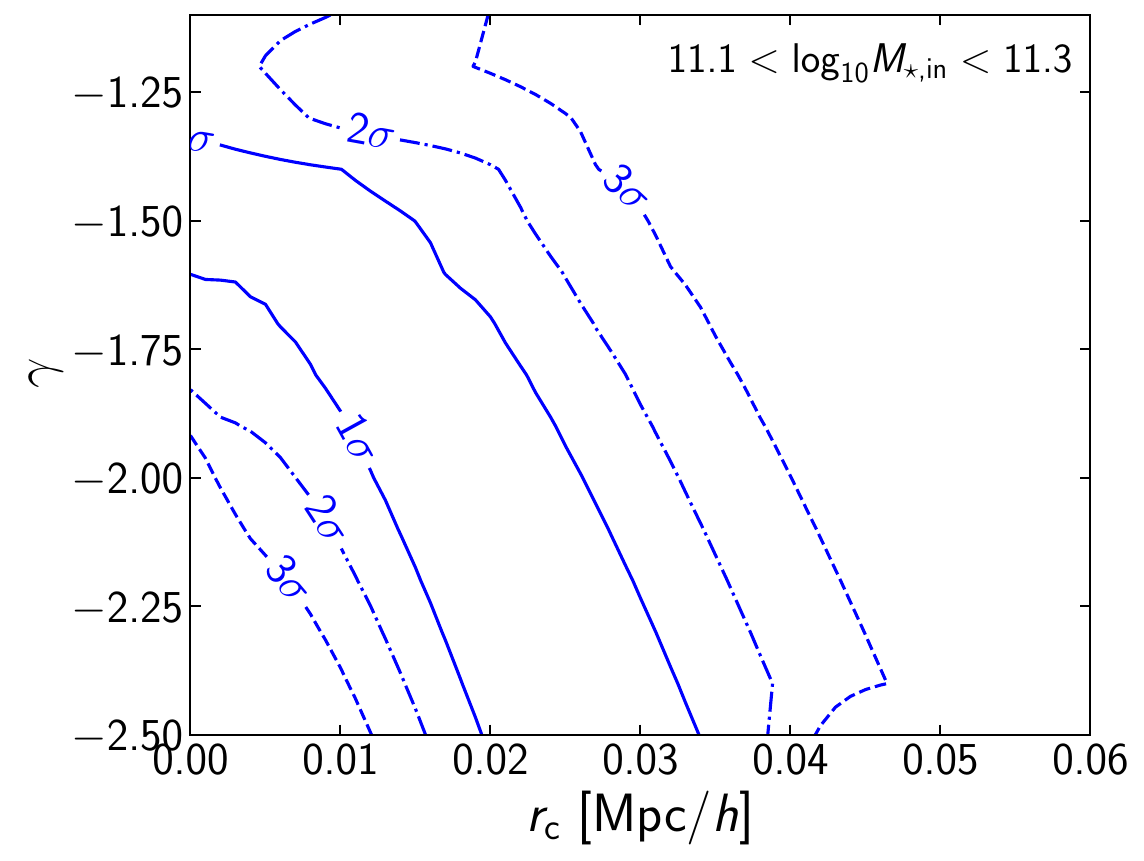}
      \includegraphics[width=0.245\linewidth]{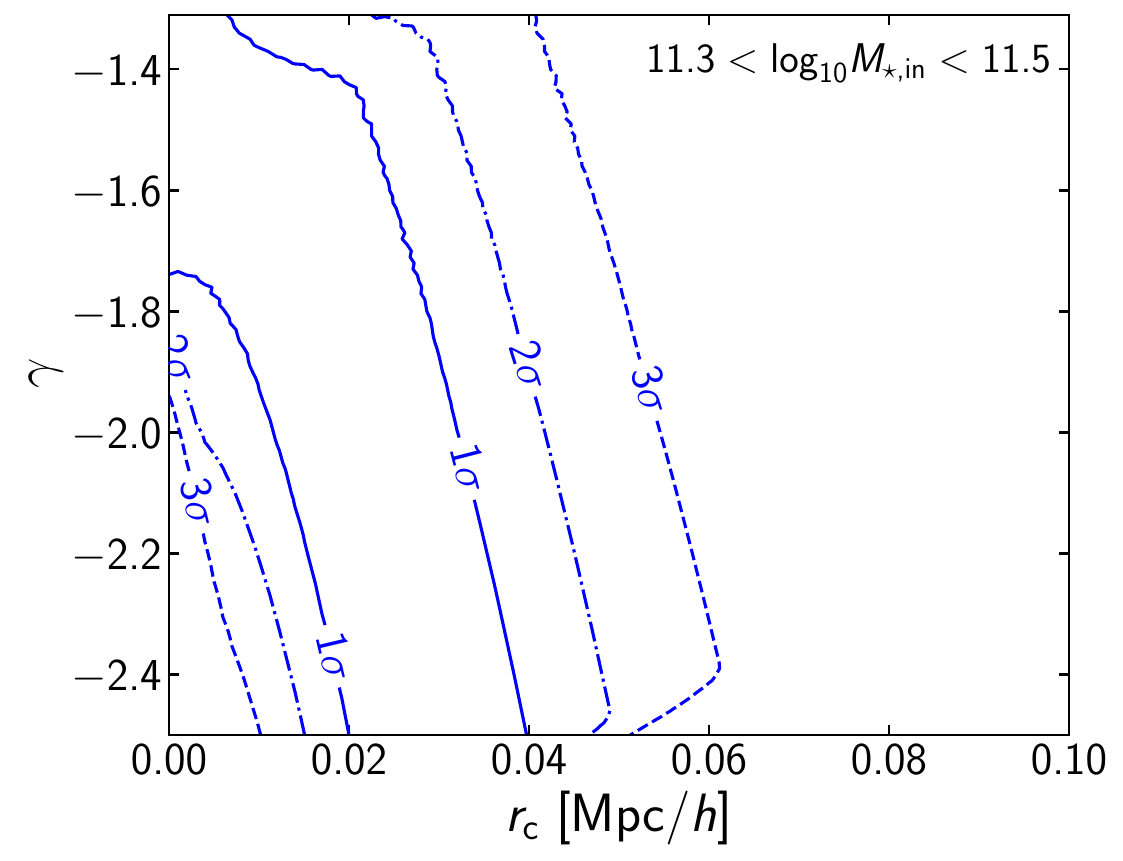}
      \includegraphics[width=0.245\linewidth]{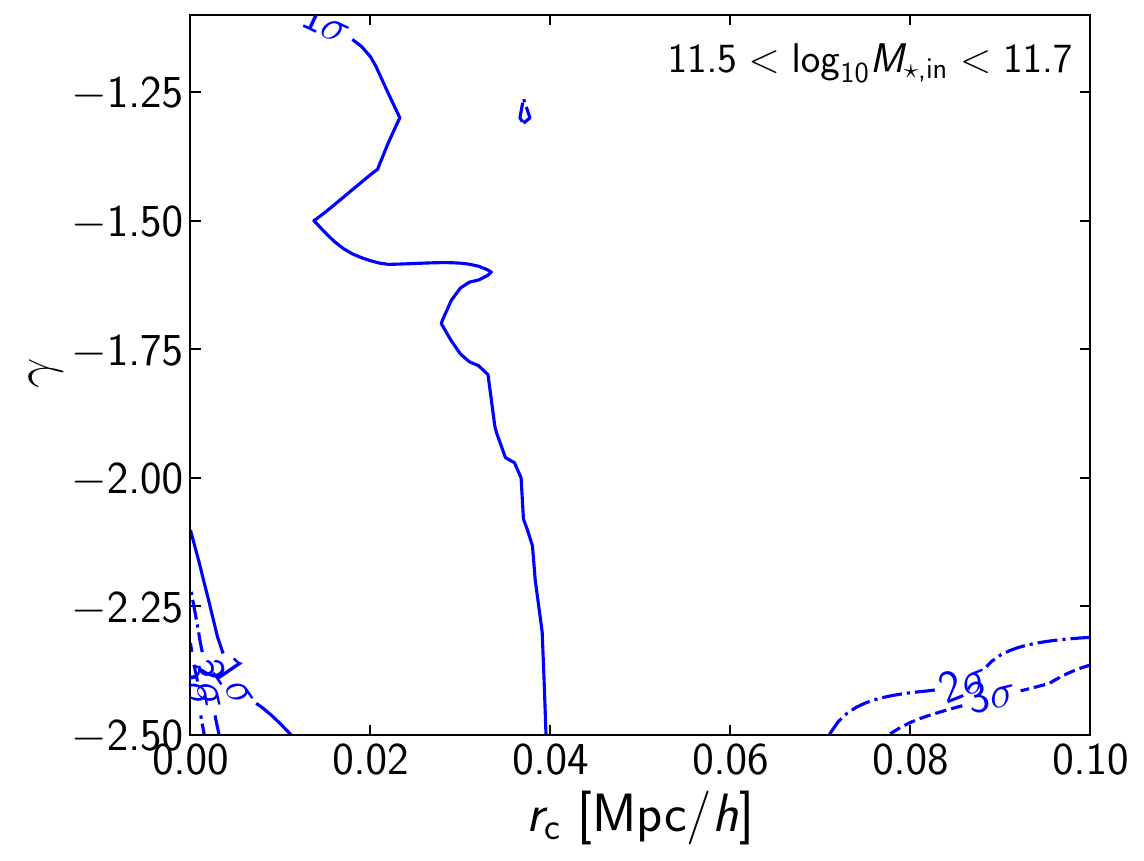}
  \end{flushleft}

  \caption{The projected constraint on parameters from the inner profile fitting in the $\gamma$-$r_{\mathrm{c}}$ plane. Different panels show results for different stellar mass bins.}
  \label{fig:gamma}
\end{figure*}

\section{Result}
\label{sec:result}

\subsection{Inner profile fitting}
\label{subsec:inner}

In Figure~\ref{fig:inner fit}, we show the differential surface density profiles obtained from the weak lensing measurements, along with the results of fitting the inner mass density profile using the model described in Equation~\eqref{eq:deltasigma}, for all stellar mass bins. 
We find that in the central region, the stellar mass component dominates, while in the outer region, the dark matter component becomes predominant. The sum of the stellar mass and dark matter components reproduces the observed differential surface density profiles well.
The best-fit parameters for each stellar mass bin are summarized in Table~\ref{table:in}.
For the stellar mass bin $10^{11.1}M_\odot<M_{\star,\mathrm{in}}<10^{11.3}M_\odot$, the best-fit $\chi^2$ value falls outside the $95\%$ confidence interval ($0.8\lesssim\chi^2\lesssim12.8$), indicating that the results for this bin should be interpreted with caution.\footnote{We note that, since we examine 35 independent $\chi^2$ including those presented in Appendix, a few $\chi^2$ values can fall outside the $95\%$ confidence interval simply by statistical flukes.  }
The result for the stellar mass bin $10^{11.3}M_\odot<M_{\star,\mathrm{in}}<10^{11.5}M_\odot$ should also be interpreted with caution, because the effect of the photometric redshift uncertainty of LRGs may not be negligible for this stellar mass bin (see Appendix~\ref{ap:spec}). 

Figure~\ref{fig:inner fit} indicates that the best-fitting profiles deviate from observed data outside the fitting range, $r>0.1~\mathrm{Mpc}/h$, which is expected because our dark matter model for fitting is an empirical (rather than physically-motivated) model. The deviations do not affect our subsequent analysis because the extrapolated profiles will not be used in any analysis.

We compare the best-fit stellar masses with those measured from the HSC-SSP photometric data.
As shown in Figure~\ref{fig:min}, the best-fit stellar masses tend to be larger than the HSC estimates, and in some stellar mass bins, the differences are significantly larger than the statistical errors. However, as noted above, the difference can be explained by the mixture of systematic errors in measurements of stellar masses in the HSC-SSP data and the uncertainty of the stellar IMF. The fact that best-fit stellar masses with weak lensing exhibit a clear positive correlation with the HSC-SSP measurements can support the validity of our lensing analysis.

Furthermore, we find a negative correlation between the radial slope of the dark matter density profile $\gamma$ and the core radius $r_\mathrm{c}$, as shown in Figure~\ref{fig:gamma}. This can be naturally explained by the fact that increasing the core radius reduces the dark matter density in the central region, which is compensated by an increase in the stellar mass. As the stellar matter component becomes more dominant, the dark matter density profile steepens, and the value of $\gamma$ decreases. Also, while core radii of dark matter distributions are consistent with zero within $2\sigma$ for most stellar mass bins, two bins with $10^{10.7}M_\odot<M_{\star,\mathrm{in}}<10^{10.9}M_\odot$ and $10^{10.9}M_\odot<M_{\star,\mathrm{in}}<10^{11.1}M_\odot$ show core radii that remain inconsistent with zero even within $3\sigma$, suggesting that cored dark matter density profiles are observed for these galaxies.

We caution that core-like profiles inferred from stacked weak gravitational lensing do not necessarily indicate physically core-like density profiles, but instead might be produced by the offset of galaxy positions from halo centers, which is sometimes referred to as wobbling. While such wobbling is expected to be small in the standard CDM model, typically much smaller than the inner boundary of our inner profile fitting, it can be larger in some dark matter scenarios such as the self-interacting dark matter model \citep{2019MNRAS.488.1572H}. The analysis of strong lens systems also indicates that such offset is typically quite small, $<1\,\mathrm{kpc}/h$ \citep[e.g.,][]{2020OJAp....3E..10W}. Nevertheless, the fact that stacked weak lensing alone cannot discriminate these two possibilities should be kept in mind when interpreting the result.

\begingroup 
    \setlength{\tabcolsep}{10pt} 
    \renewcommand{\arraystretch}{1.5} 
    \begin{table*}[t]
        \centering
        \begin{tabular}{ ccccc }  
    \hline \hline
$\log_{10}(M_{\star,\mathrm{in}}[M_\odot])$ &
$M_{\mathrm{NFW}}$ [$M_\odot/h$] &
$M_{\mathrm{h}}$ [$M_\odot/h$] &
$f_\mathrm{sat}$ &
$\chi^2_\mathrm{out}/\mathrm{dof}$ \\
            \hline     
10.3--10.5 & $1.2^{+0.1}_{-0.1} \times 10^{12}$ & $5.1^{+0.1}_{-0.1} \times 10^{13}$ & $0.40^{+0.00}_{-0.01}$ & 19.3/16 \\
            \hline
10.5--10.7 & $1.3^{+0.1}_{-0.1} \times 10^{12}$ & $5.4^{+0.5}_{-0.4} \times 10^{13}$ & $0.32^{+0.02}_{-0.02}$ & 20.9/16 \\
            \hline
10.7--10.9 & $1.4^{+0.1}_{-0.1} \times 10^{12}$ & $5.5^{+0.6}_{-0.1} \times 10^{13}$ & $0.27^{+0.02}_{-0.02}$ & 19.3/16 \\
            \hline
10.9--11.1 & $1.8^{+0.1}_{-0.1} \times 10^{12}$ & $6.8^{+0.9}_{-0.7} \times 10^{13}$ & $0.21^{+0.02}_{-0.02}$ & 31.4/16 \\
            \hline
11.1--11.3 & $2.6^{+0.2}_{-0.1} \times 10^{12}$ & $6.7^{+1.1}_{-0.6} \times 10^{13}$ & $0.21^{+0.02}_{-0.03}$ & 13.31/16 \\
            \hline
11.3--11.5 & $4.7^{+0.4}_{-0.3} \times 10^{12}$ & $1.1^{+0.4}_{-0.3} \times 10^{14}$ & $0.14^{+0.03}_{-0.03}$ & 19.2/16 \\
            \hline
11.5--11.7 & $7.6^{+1.3}_{-1.2} \times 10^{12}$ & $4.9^{+4.1}_{-1.6} \times 10^{13}$ & $0.30^{+0.10}_{-0.10}$ & 16.0/16 \\
            \hline \hline  
        \end{tabular}  
        \caption{Parameters used in the outer profile fitting. Here, $M_{\mathrm{NFW}}$ corresponds to $M$ in Section~\ref{subsec:outer model}. The errors are shown at the $1\sigma$ level.}
        \label{table:out}
    \end{table*}
\endgroup

\begin{figure*}[tbp]
  \centering
    \includegraphics[width=0.245\linewidth]{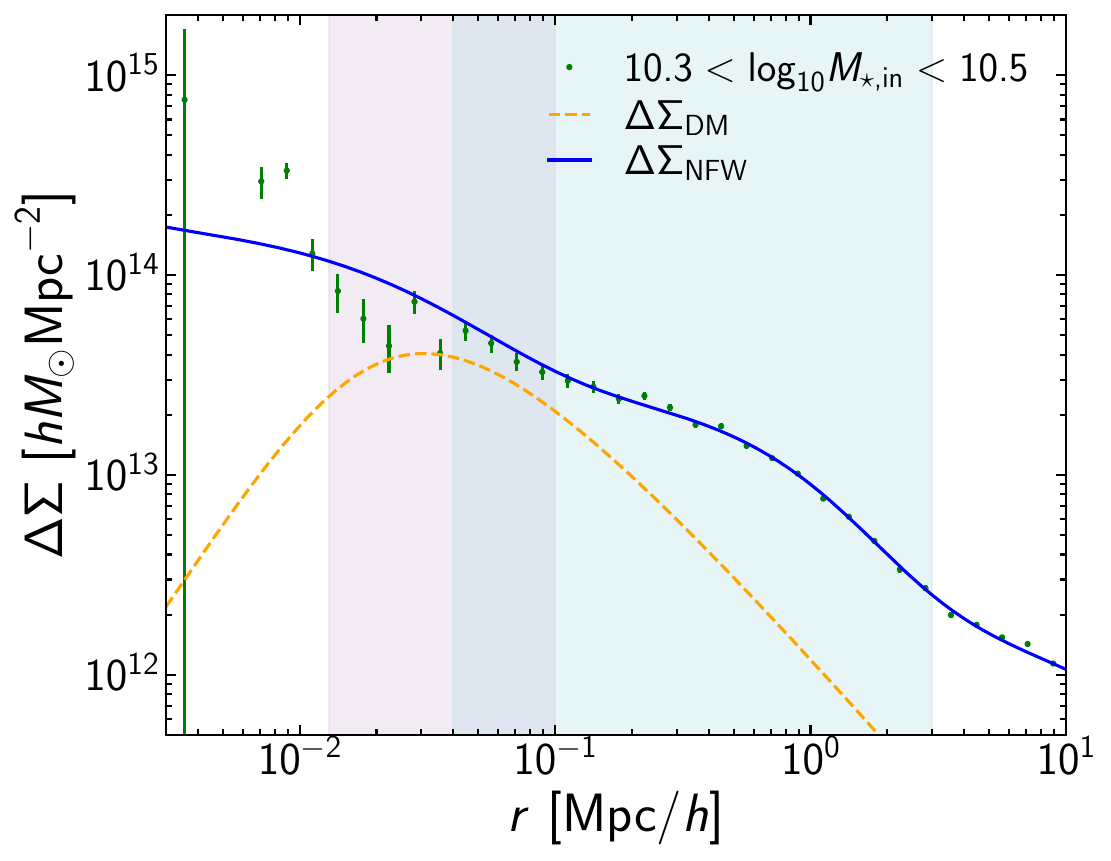}
    \includegraphics[width=0.245\linewidth]{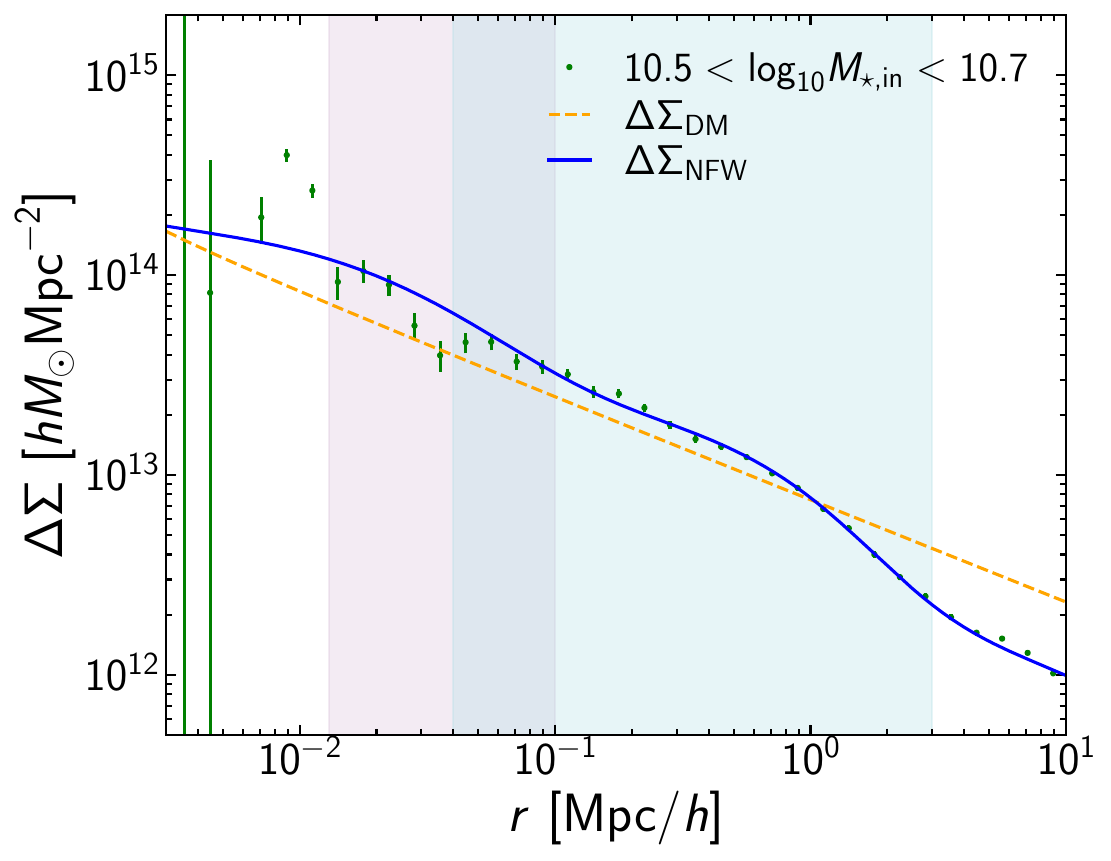}
    \includegraphics[width=0.245\linewidth]{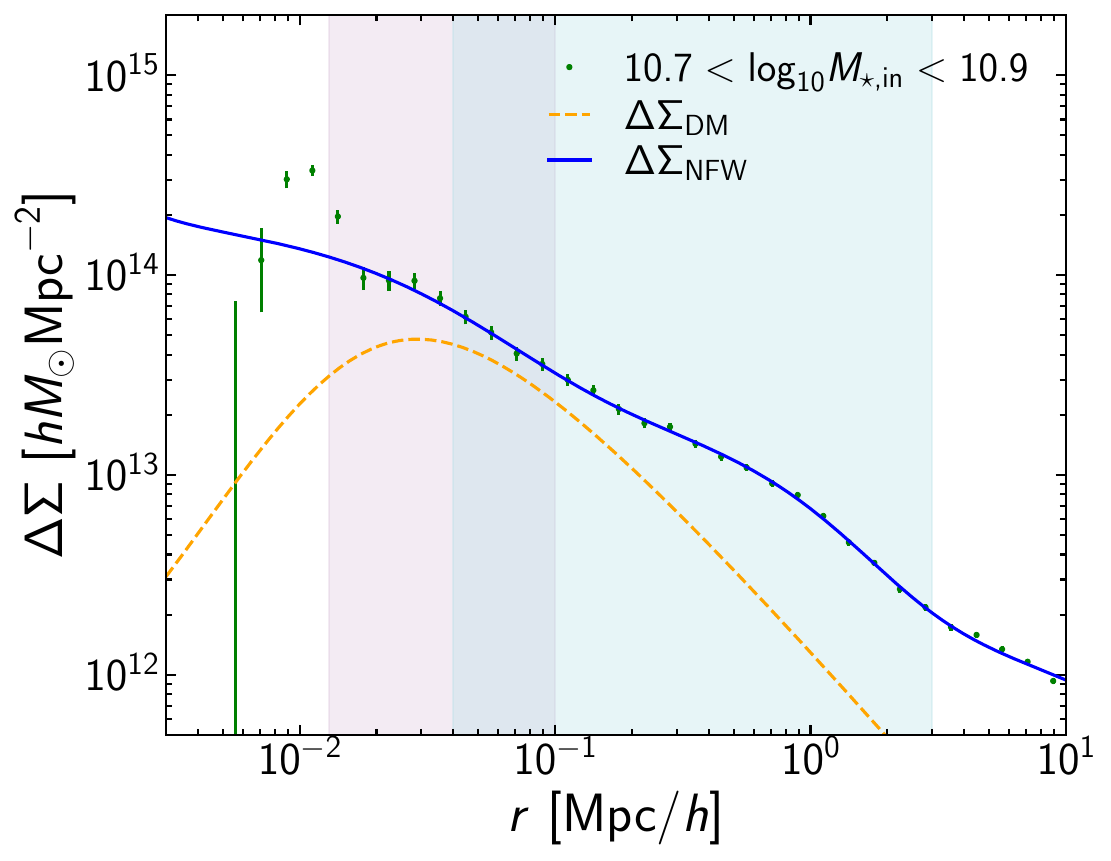}
    \includegraphics[width=0.245\linewidth]{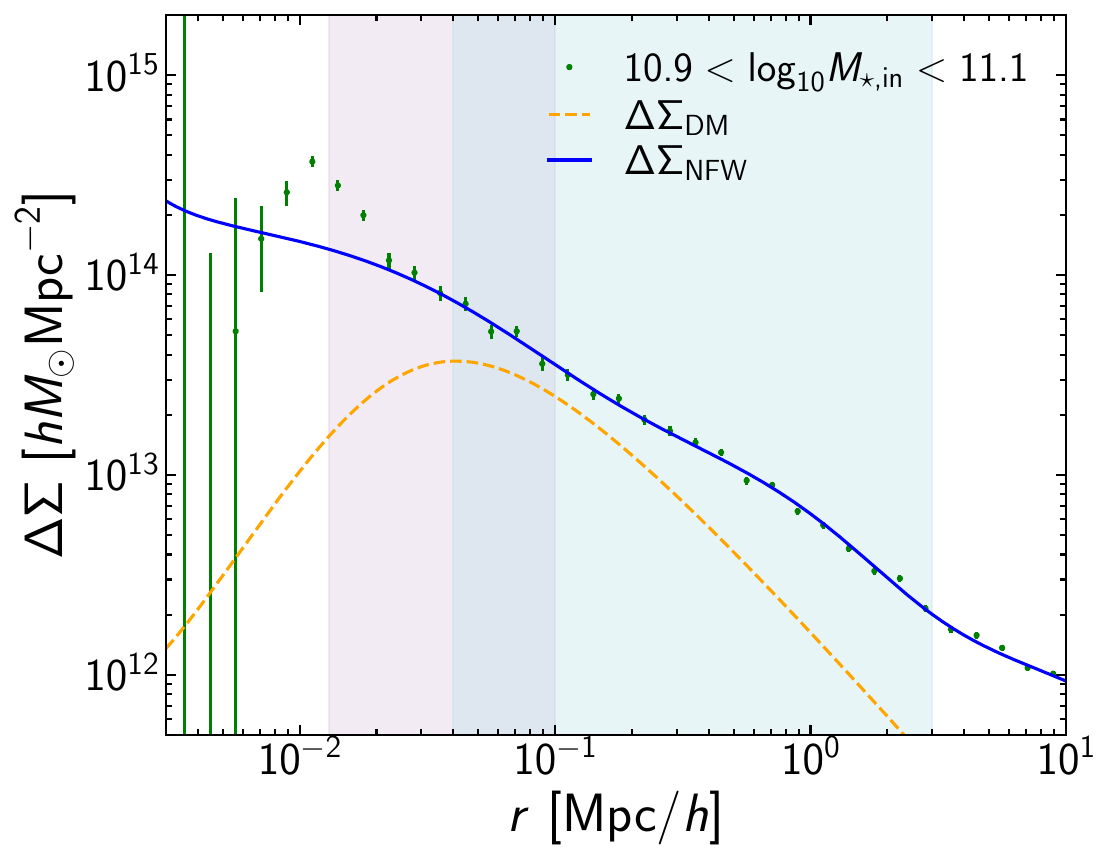}

  \vskip\baselineskip

  \begin{flushleft}
      \includegraphics[width=0.245\linewidth]{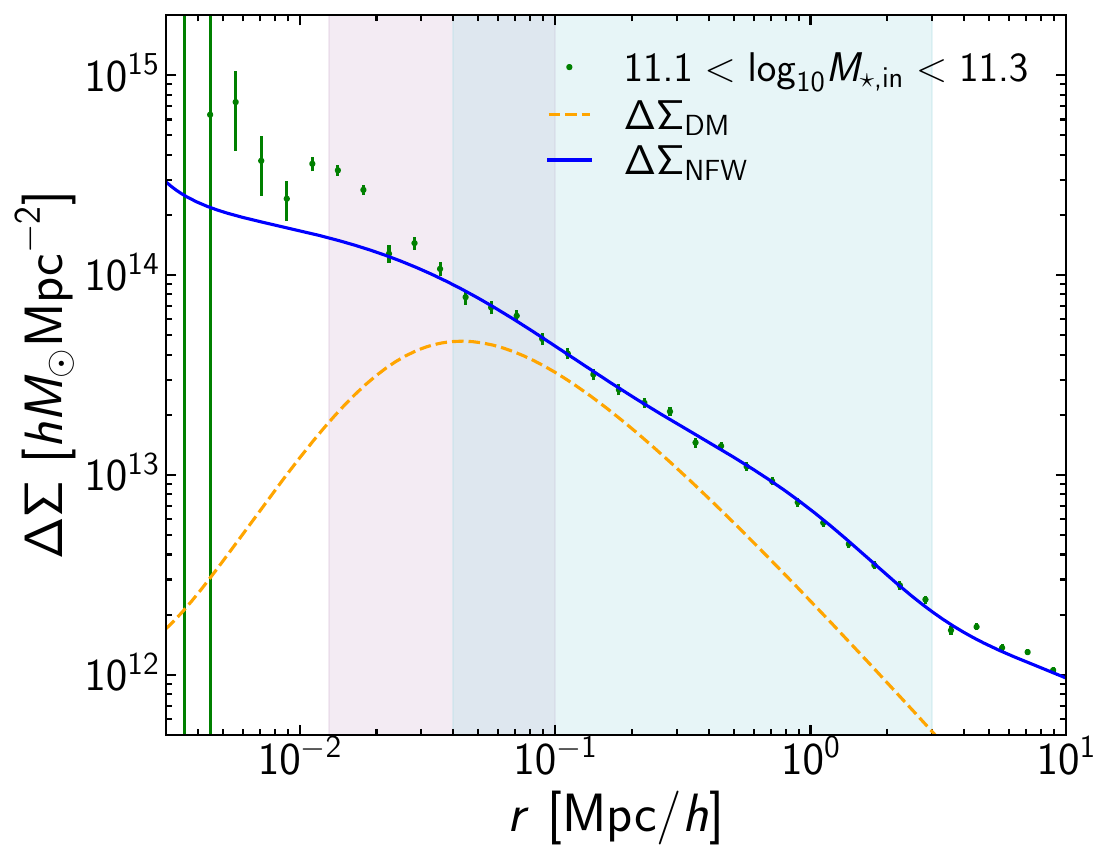}
      \includegraphics[width=0.245\linewidth]{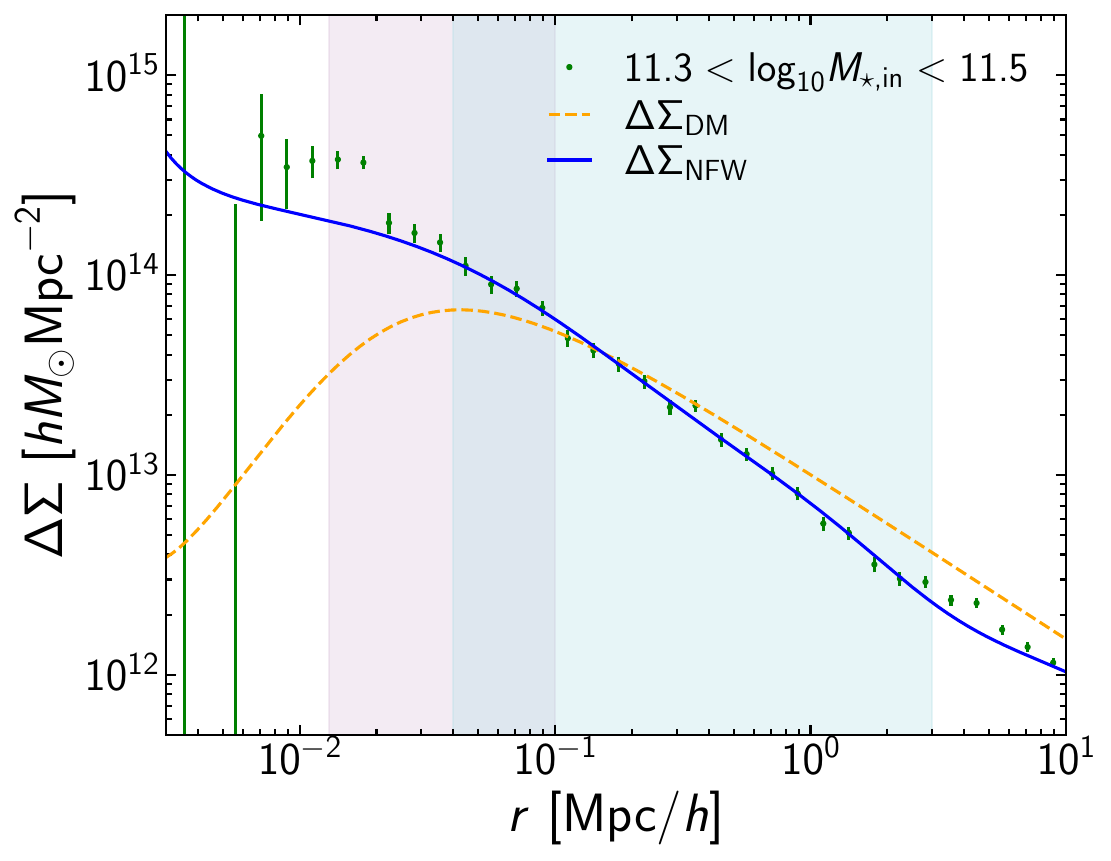}
      \includegraphics[width=0.245\linewidth]{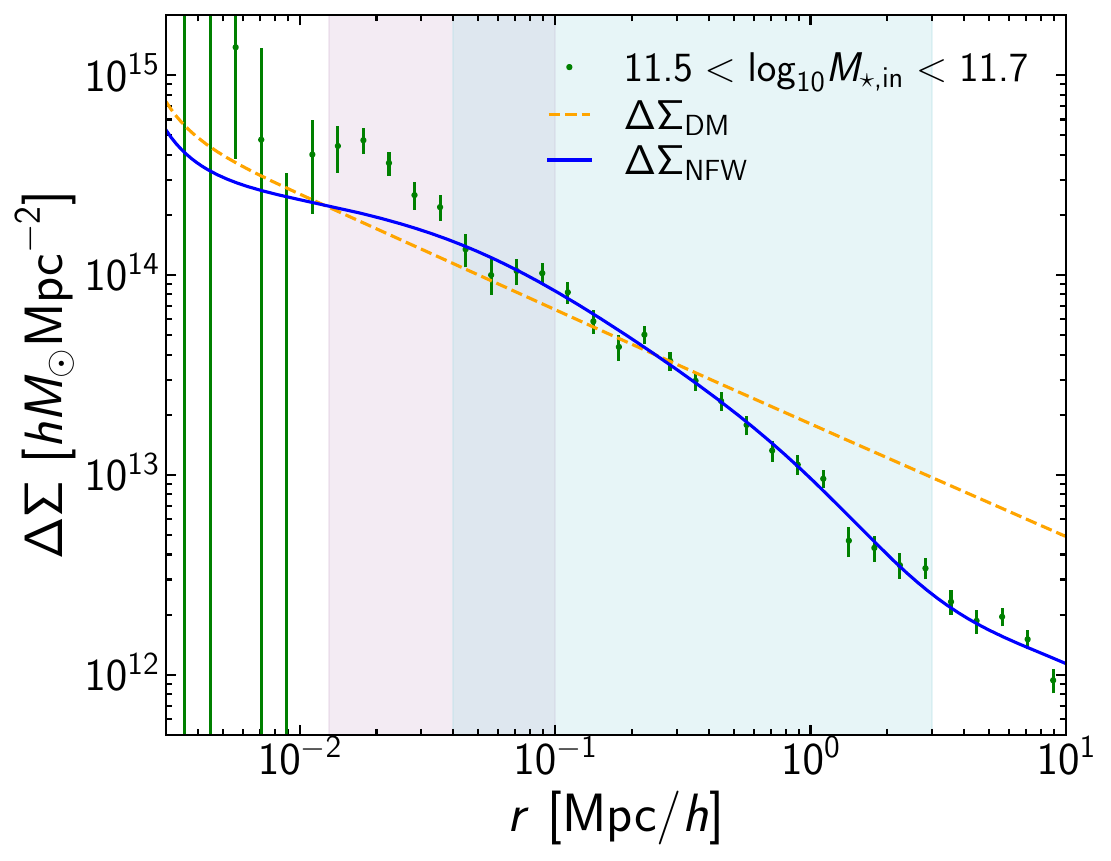}
  \end{flushleft}

  \caption{Comparison between the dark matter density profile obtained from the inner profile fitting and the NFW profile fitted to the outer lensing profile, where the NFW profile includes additional contributions from the satellite component as well as 2-halo term (see Equation~\ref{eq:NFW}). Different panels show results for different stellar mass bins. The orange dashed and blue solid lines indicate the dark matter distribution inferred from inner profile fitting and the NFW profile from the outer profile fitting, respectively. The magenta-shaded region denotes the inner-fitting range, and the light-blue shaded region shows the outer-fitting range.}
  \label{fig:nfw}
\end{figure*}

\begin{figure*}[tbp]
  \centering
    \includegraphics[width=0.245\linewidth]{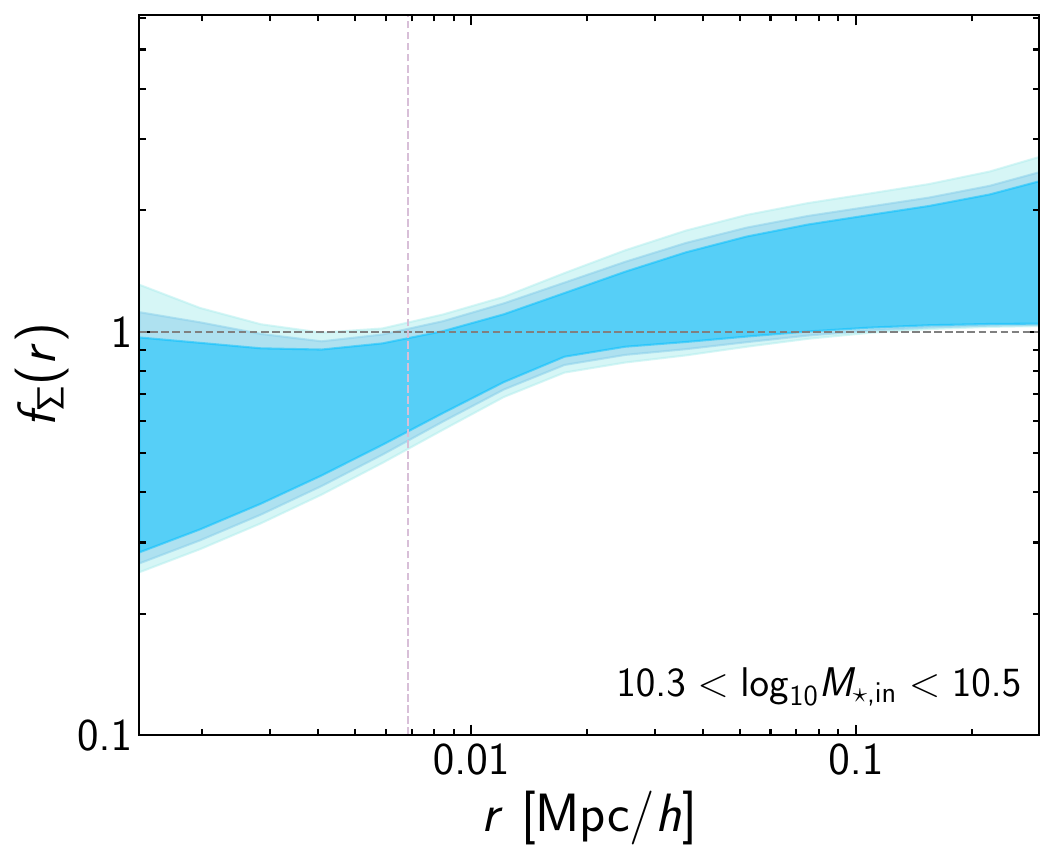}
    \includegraphics[width=0.245\linewidth]{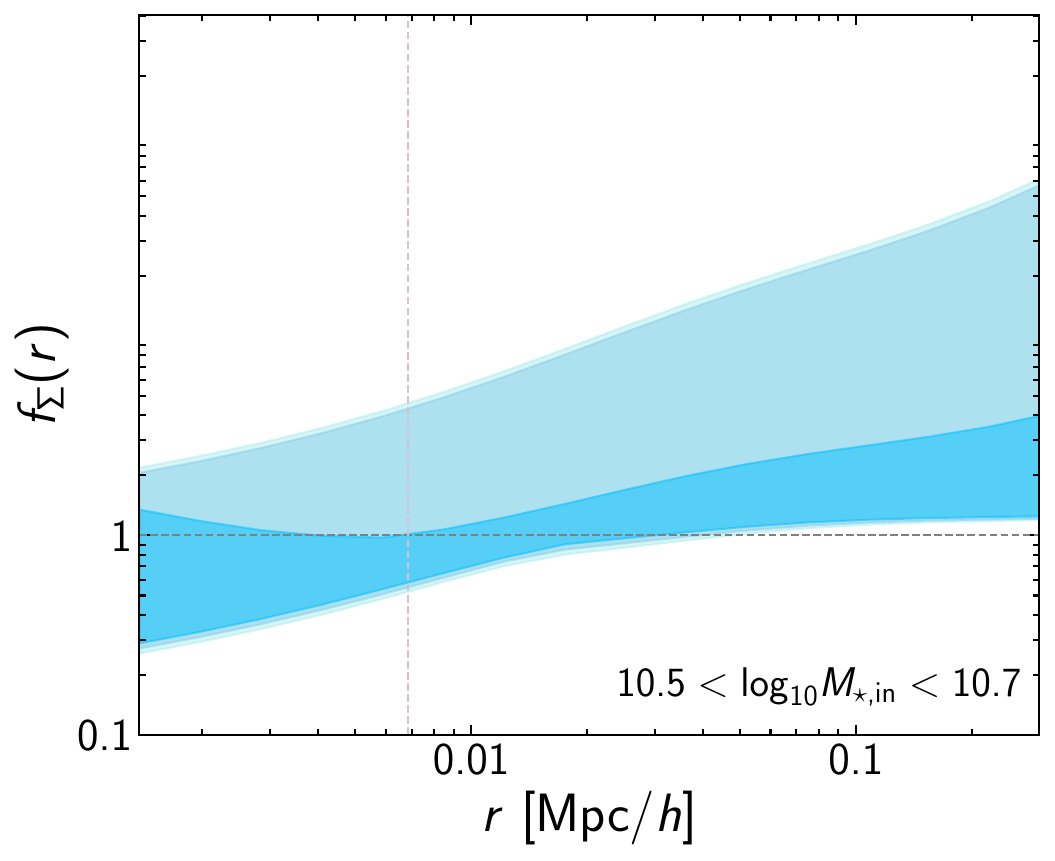}
    \includegraphics[width=0.245\linewidth]{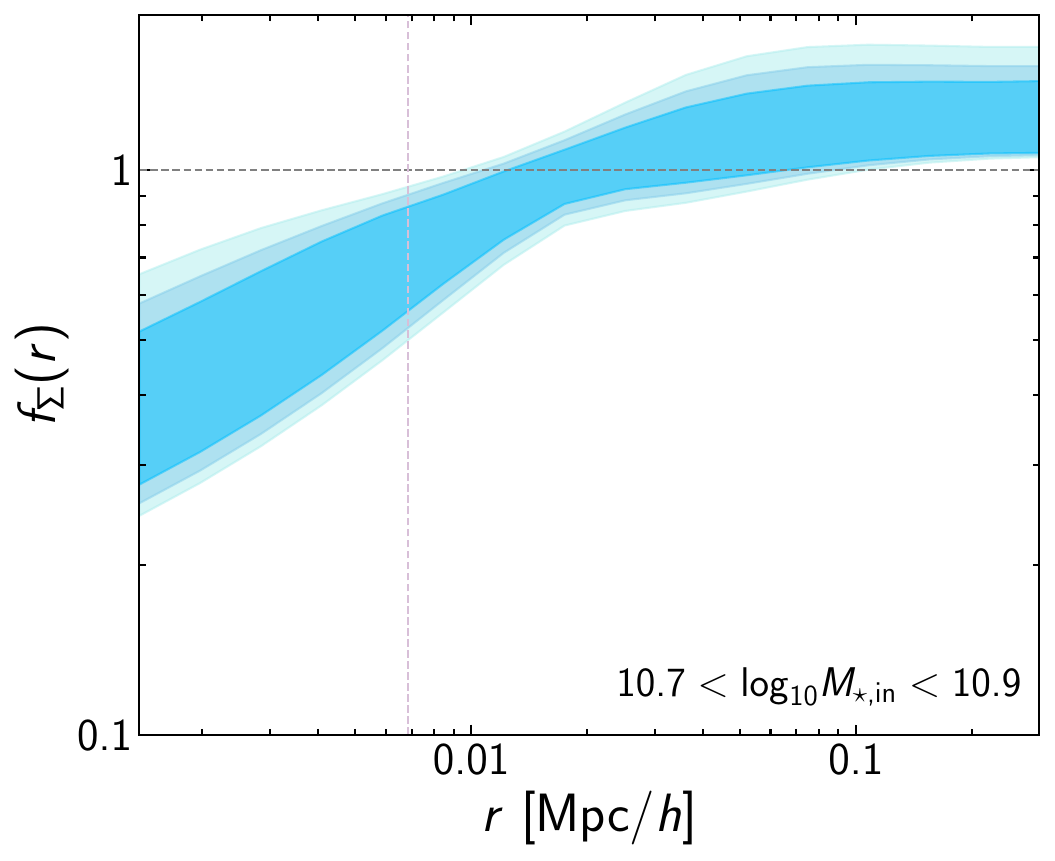}
    \includegraphics[width=0.245\linewidth]{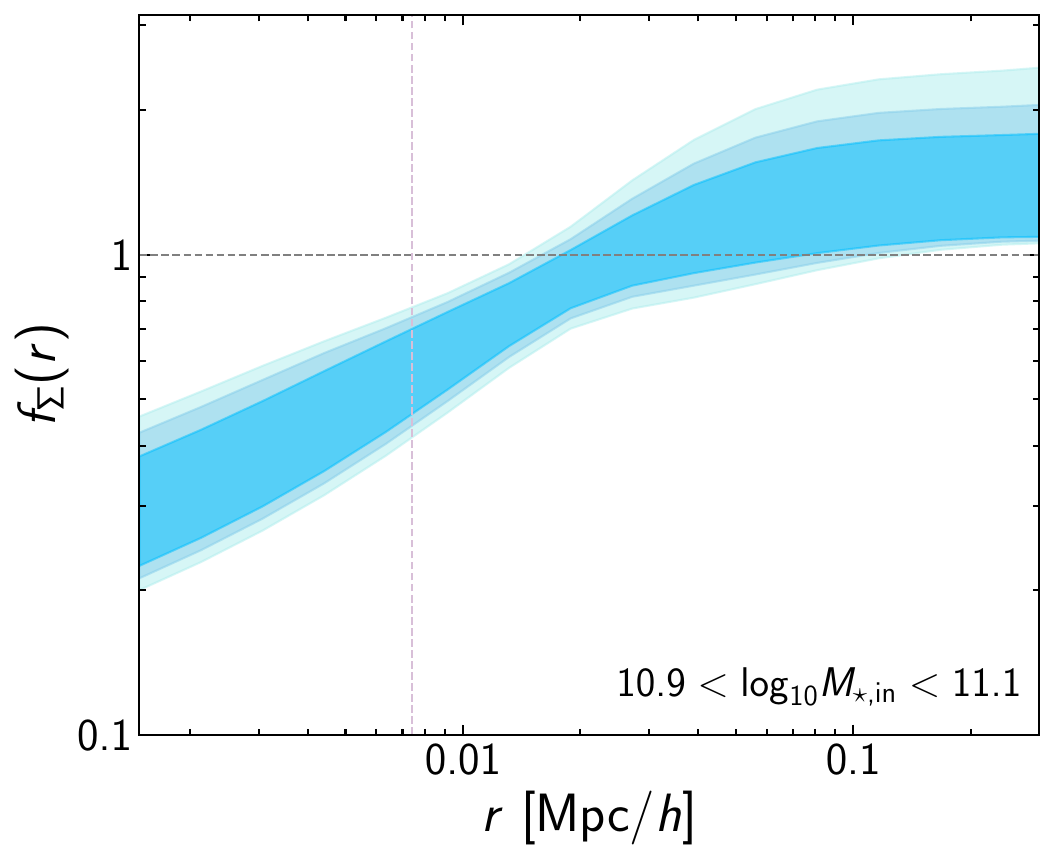}

  \vskip\baselineskip

  \begin{flushleft}
      \includegraphics[width=0.245\linewidth]{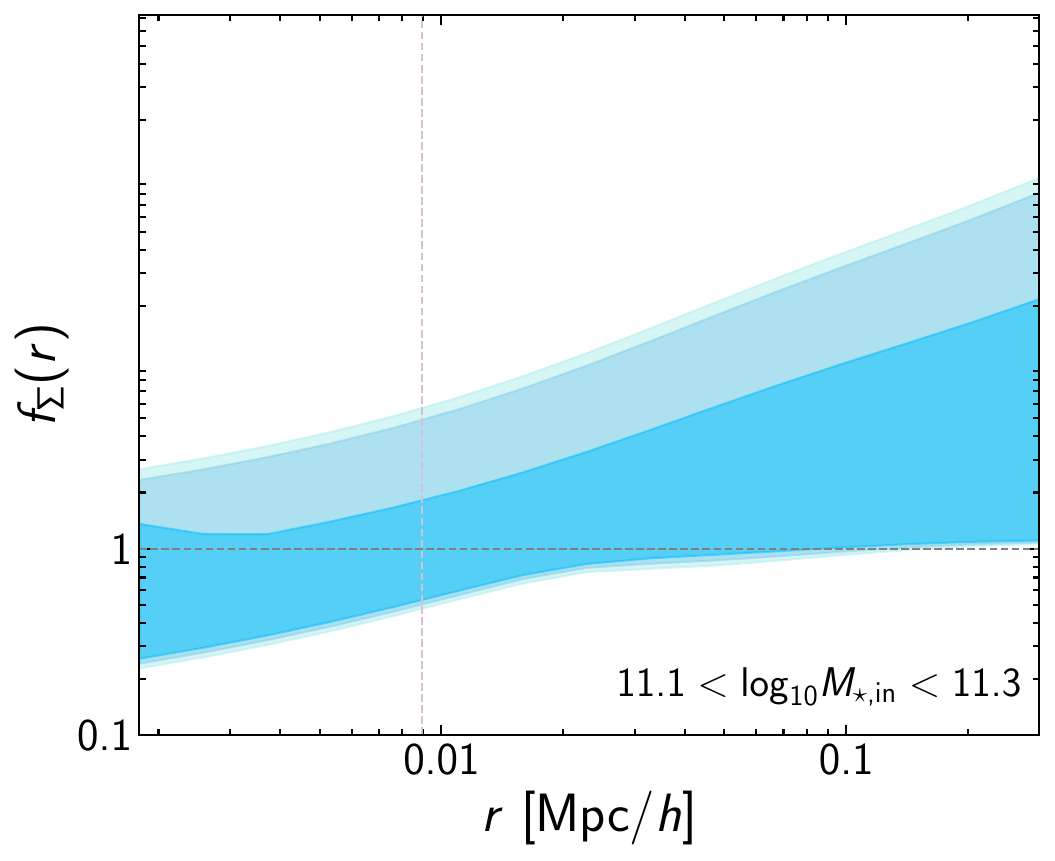}
      \includegraphics[width=0.245\linewidth]{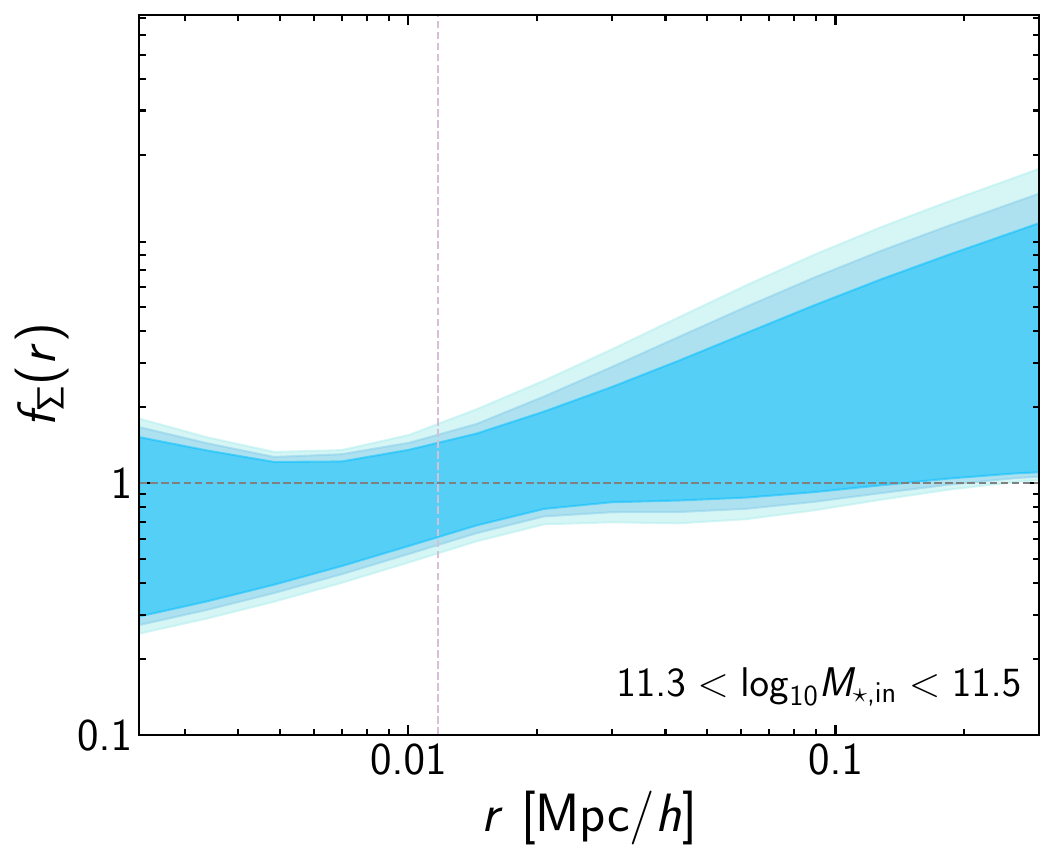}
      \includegraphics[width=0.245\linewidth]{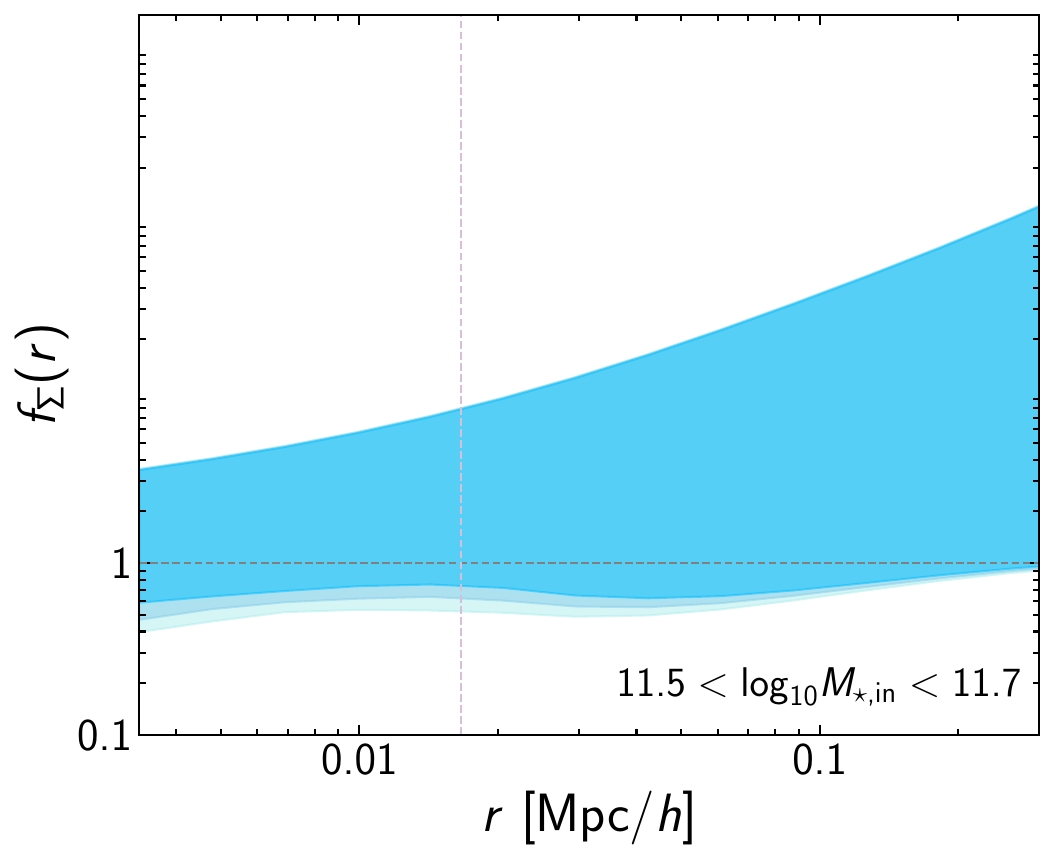}
  \end{flushleft}

  \caption{The ratio of the surface mass densities $f_{\Sigma}(r)$ defined in Equation~\eqref{eq:fsigma}. The shaded regions represent the $1\sigma$, $2\sigma$, and $3\sigma$ confidence intervals from darker to lighter colors. The vertical dashed line shows $r=5r_\mathrm{e}$. Different panels show results for different stellar mass bins.
}
  \label{fig:fsigma}
\end{figure*}

\begin{figure*}[tbp]
  \centering
    \includegraphics[width=0.245\linewidth]{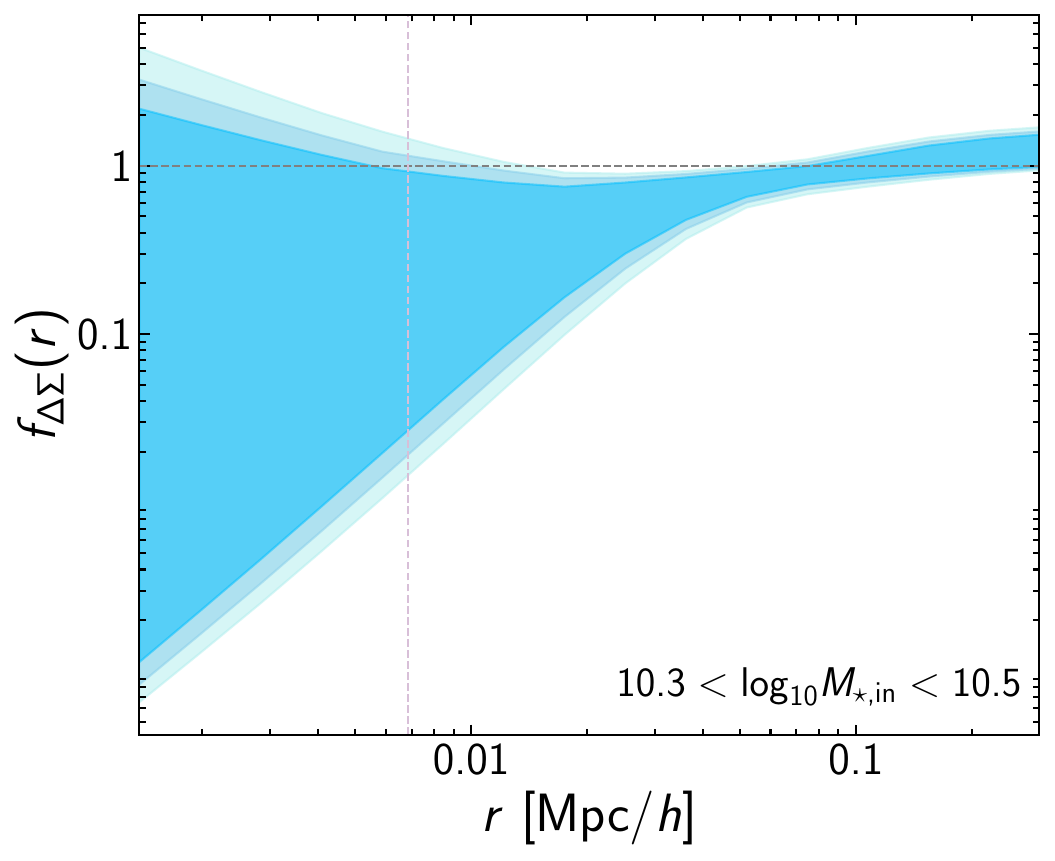}
    \includegraphics[width=0.245\linewidth]{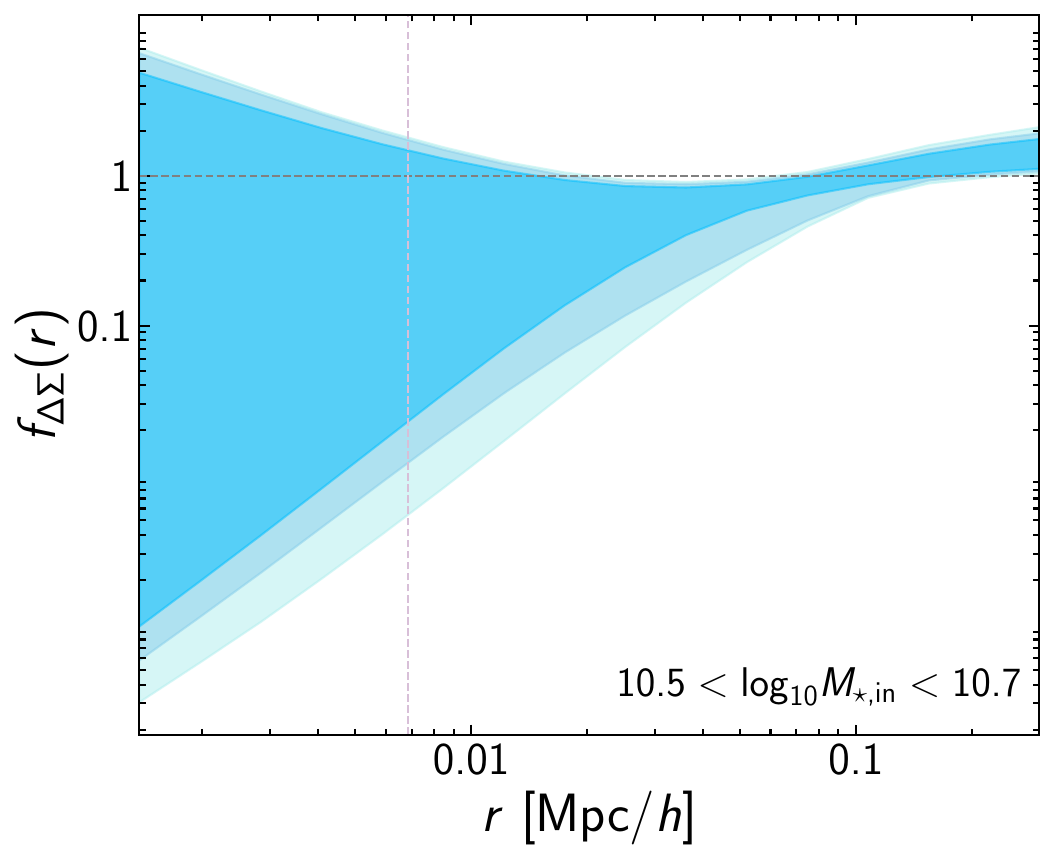}
    \includegraphics[width=0.245\linewidth]{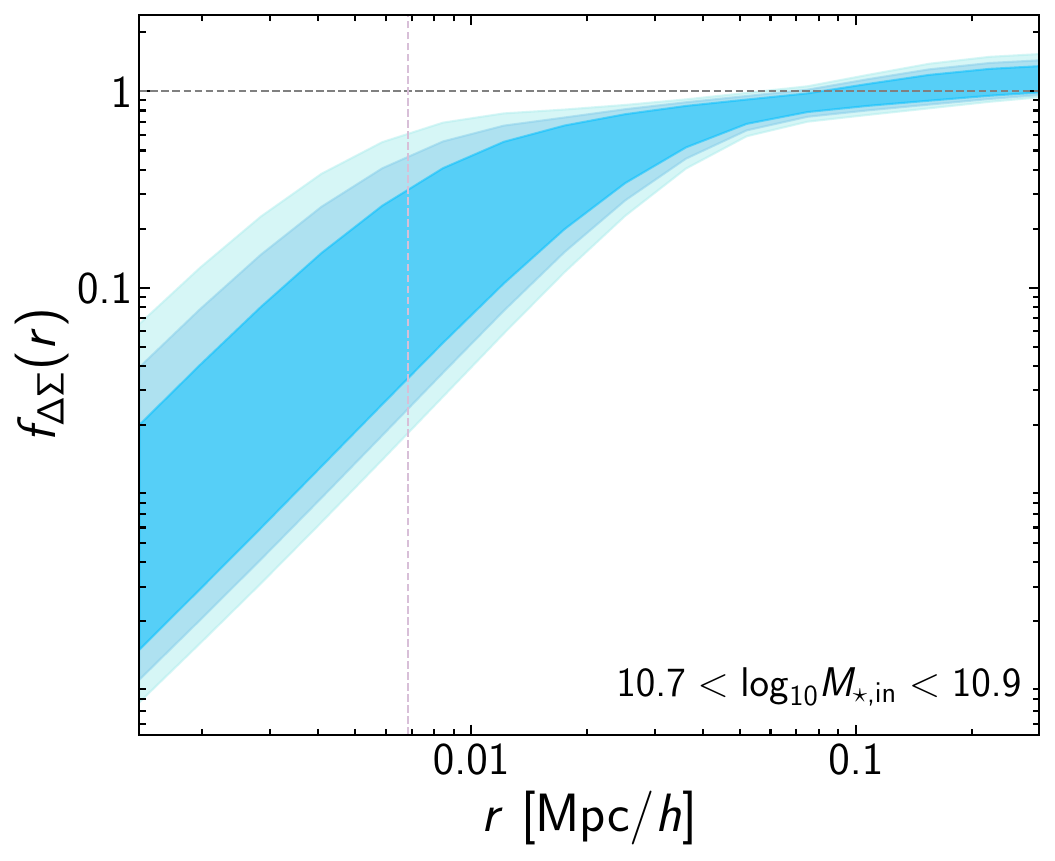}
    \includegraphics[width=0.245\linewidth]{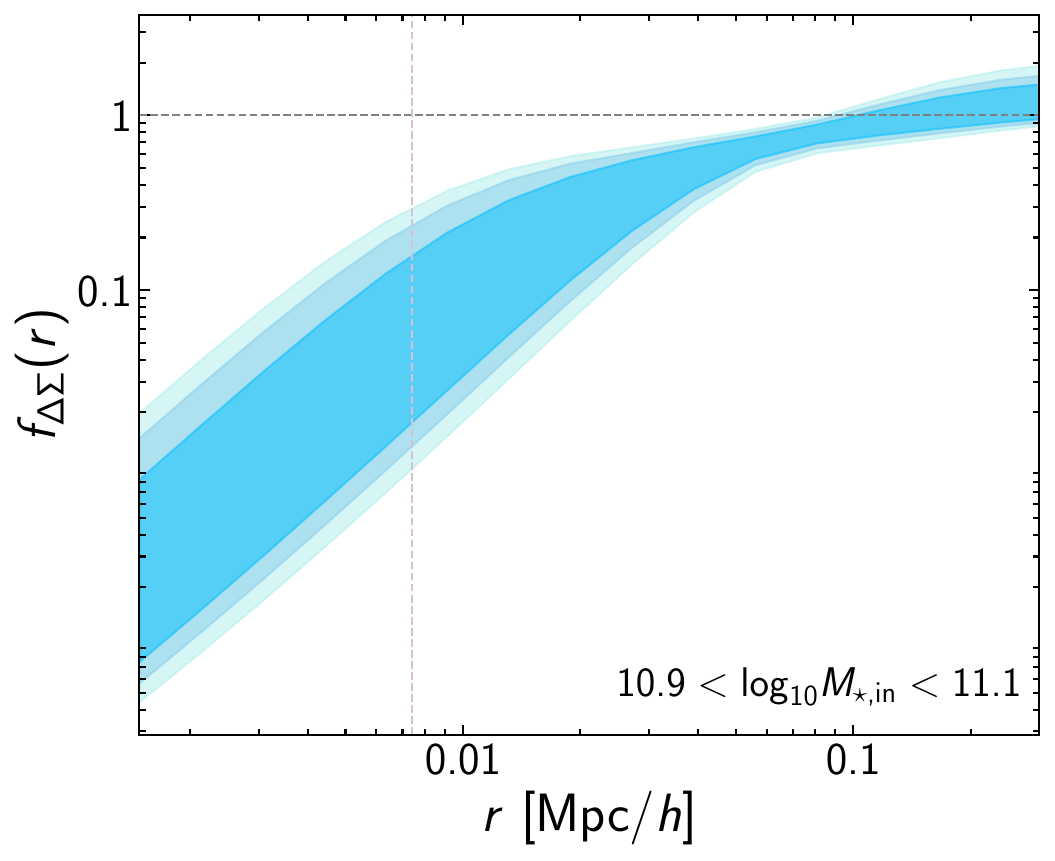}

  \vskip\baselineskip

  \begin{flushleft}
      \includegraphics[width=0.245\linewidth]{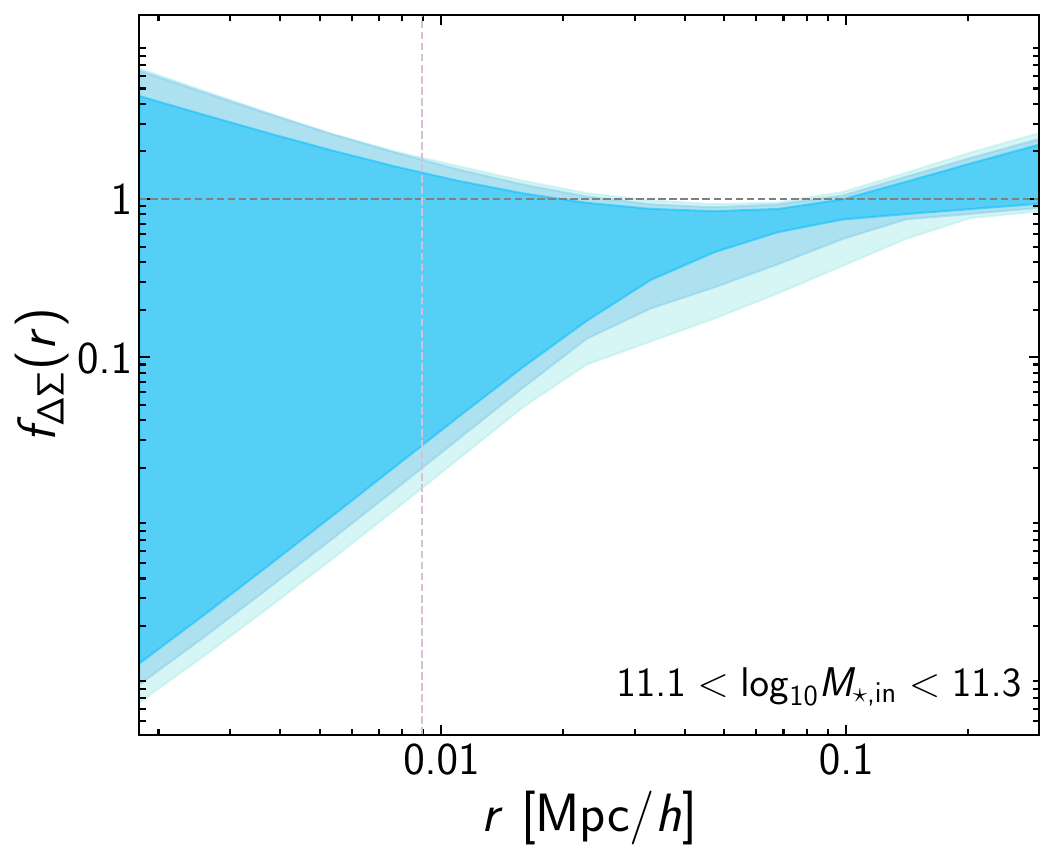}
      \includegraphics[width=0.245\linewidth]{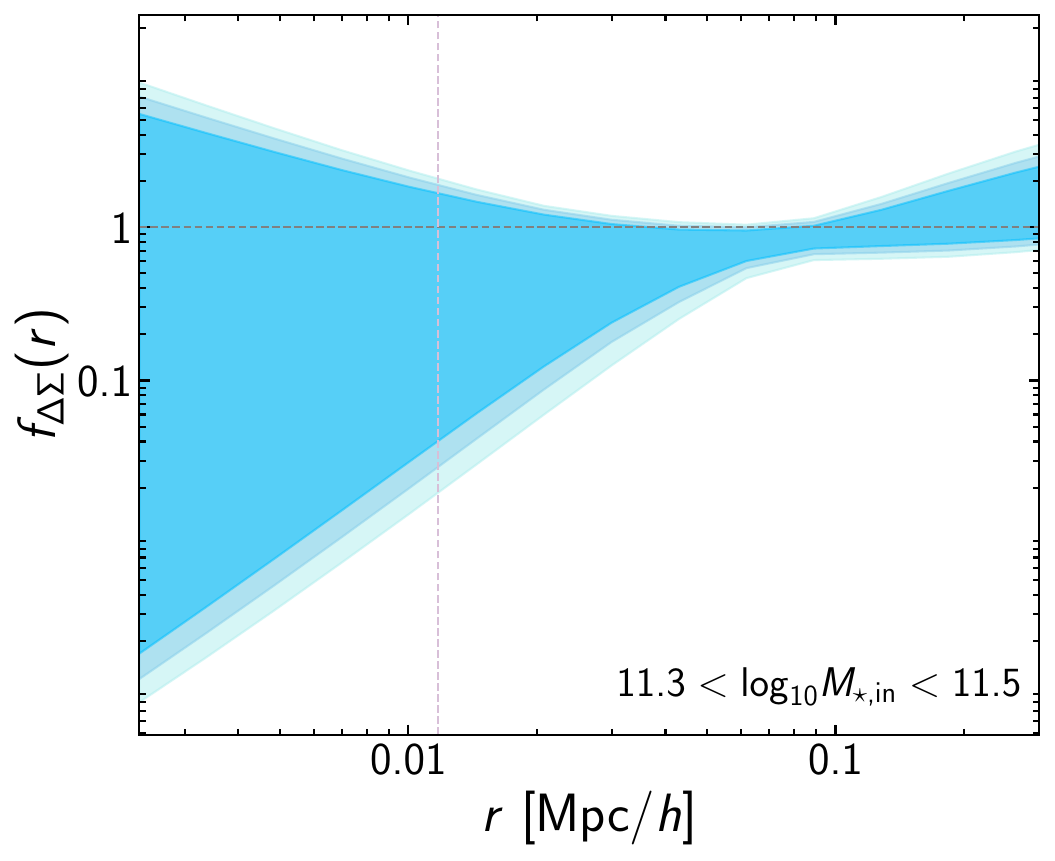}
      \includegraphics[width=0.245\linewidth]{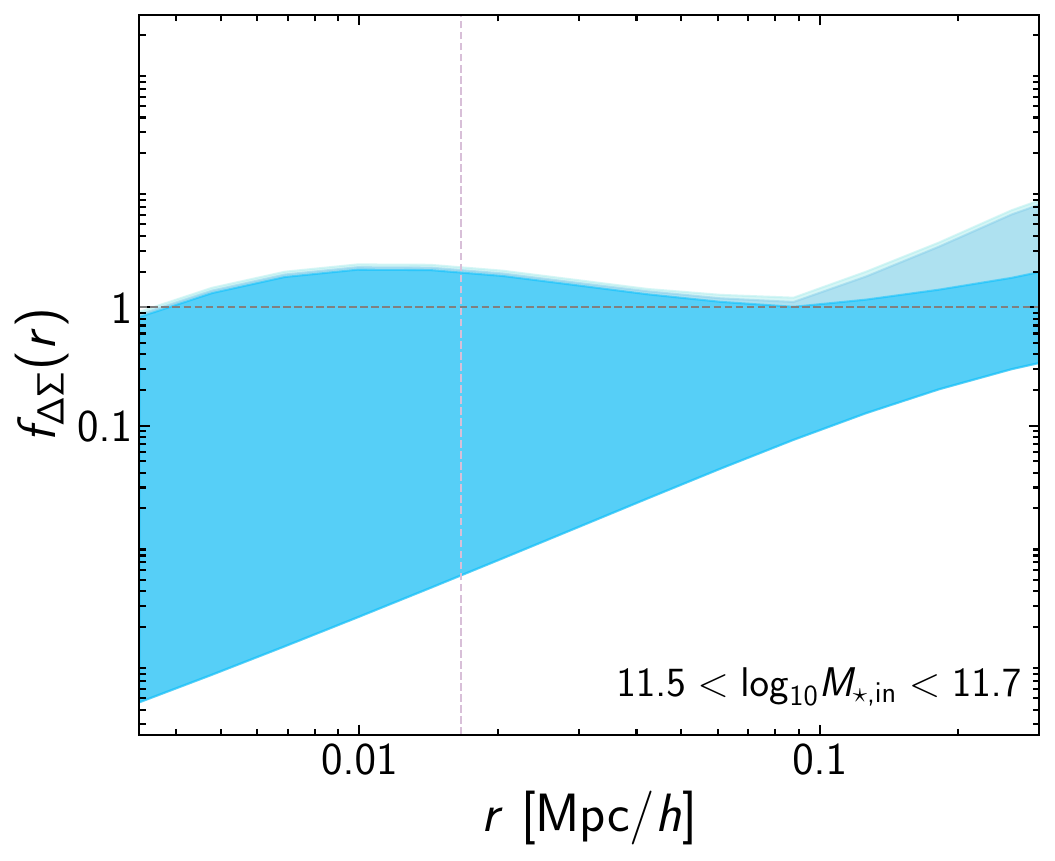}
  \end{flushleft}

  \caption{Same as Figure~\ref{fig:fsigma}, but for $f_{\Delta\Sigma}(r)$ defined in Equation~\eqref{eq:dfsigma}.}
  \label{fig:fdsigma}
\end{figure*}

\subsection{Comparison with the NFW profile}
\label{subsec:NFW}

We compare the dark matter density profile obtained from the inner profile fitting with the NFW profile. For this purpose, the outer lensing profile is fitted with an NFW-based model given by Equation~\eqref{eq:NFW} to determine the mass of the central NFW component. The best-fit parameters for each stellar mass bin are summarized in Table~\ref{table:out}. The resulting NFW profile is then extrapolated inward and compared with the dark matter distribution obtained from the inner profile fitting. The results of the outer profile fitting are shown in Figure~\ref{fig:nfw}.

For a more quantitative comparison, we evaluate their differences with the ratio of the two profiles.
Specifically, we define
\begin{equation}
  f_{\Sigma}(r) = \frac{\Sigma_{\mathrm{DM}}(r)}{\Sigma_{\mathrm{NFW}}(r)},
  \label{eq:fsigma}
\end{equation}
as the ratio of the surface mass densities and 
\begin{equation}
  f_{\Delta\Sigma}(r) = \frac{\Delta\Sigma_{\mathrm{DM}}(r)}{\Delta\Sigma_{\mathrm{NFW}}(r)},
  \label{eq:dfsigma}
\end{equation}
as the ratio of the differential surface mass densities. We show $f_{\Sigma}(r)$ and $f_{\Delta\Sigma}(r)$ in Figures~\ref{fig:fsigma} and ~\ref{fig:fdsigma}, respectively.
In the two stellar mass bins where cored dark matter distributions are identified, both $f_{\Sigma}(r)$ and $f_{\Delta\Sigma}(r)$ fall below $1$ in the central region, and are inconsistent with $1$ at more than $3\sigma$ level. This result supports the interpretation that the central dark matter profiles in these two stellar mass bins deviate from the NFW form. 
For the other stellar mass bins, $f_{\Sigma}(r)$ and $f_{\Delta\Sigma}(r)$ are consistent with $1$ within $\sim2\sigma$, indicating that the central dark matter profiles agree well with the NFW profile. 

The dark matter distribution at centers of early-type galaxies is also extensively studied by strong gravitational lensing. A caveat is that stellar masses of lensing galaxies of strong lens systems are typically $\sim 3\times 10^{11}M_\odot$ assuming the Salpeter IMF \cite[see Figure 1 of][]{2014MNRAS.439.2494O}, which corresponds to the high mass end of our stellar mass bins where the central dark matter profiles are found to be consistent with the NFW profile. In agreement with several previous strong lensing studies \citep[e.g.,][]{2010ApJ...721L.163A,2011MNRAS.416..322D,2013ApJ...765...25N,2014MNRAS.439.2494O,2021MNRAS.503.2380S,2025MNRAS.541....1S}, our result appears to disfavor the enhancement of central dark matter profiles due to the adiabatic contraction \citep[e.g.,][]{2004ApJ...616...16G}.

\begingroup 
    \setlength{\tabcolsep}{10pt} 
    \renewcommand{\arraystretch}{1.5} 
    \begin{table}[t]
        \centering
        \begin{tabular}{ cc }  
    \hline \hline
$\log_{10}(M_{\star,\mathrm{in}}[M_\odot])$ & $f_{\mathrm{DM}}(<5r_\mathrm{e})$ \\
            \hline     
10.3--10.5 & $0.19^{+0.70}_{-0.12}$ \\
            \hline
10.5--10.7 & $0.51^{+0.45}_{-0.47}$ \\
            \hline
10.7--10.9 & $0.087^{+0.125}_{-0.050}$ \\
            \hline
10.9--11.1 & $0.035^{+0.050}_{-0.017}$ \\
            \hline
11.1--11.3 & $0.056^{+0.431}_{-0.025}$ \\
            \hline
11.3--11.5 & $0.14^{+0.53}_{-0.08}$ \\
            \hline
11.5--11.7 & $0.50^{+0.42}_{-0.49}$ \\
            \hline \hline  
        \end{tabular}  
        \caption{The values of the dark matter fractions $f_{\mathrm{DM}}(<5r_\mathrm{e})$. The errors are shown at the $1\sigma$ level.}
        \label{table:fdm}
    \end{table}
\endgroup

\begin{figure}[tbp]
  \includegraphics[width=\linewidth]{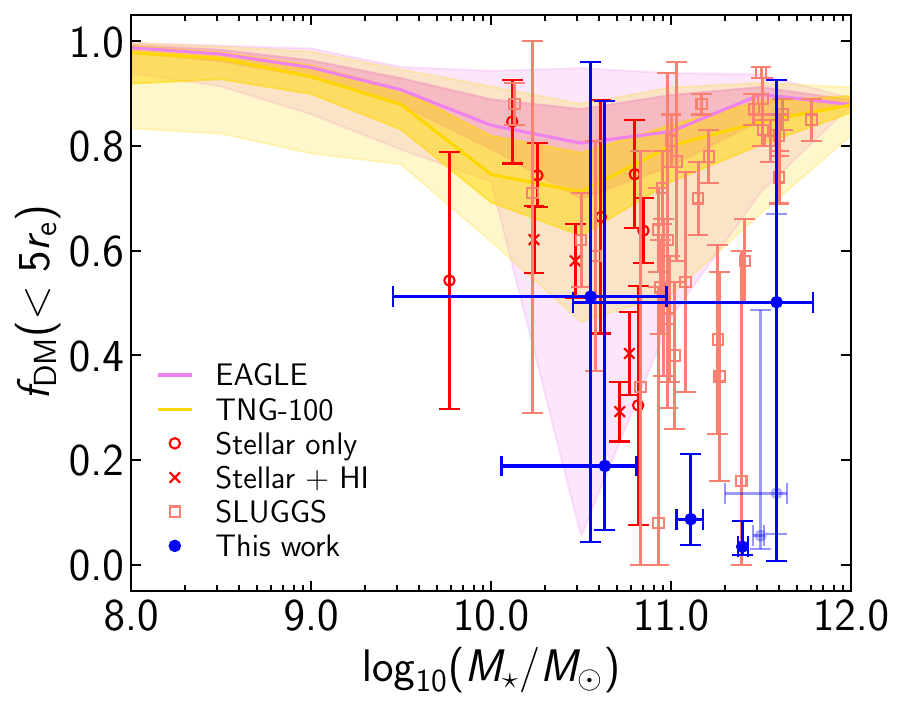}
  \caption{Dark matter fractions $f_{\mathrm{DM}}(<5r_\mathrm{e})$. The blue points show our results. Open circles and crosses represent the results obtained from the orbit-based dynamical models \citep{2024MNRAS.528.5295Y}, and open squares indicate the results from the SLUGGS survey \citep{2017MNRAS.468.3949A}.} For comparison, solid violet and gold curves shows the predictions of the EAGLES and TNG-100 simulations, respectively, with dark and light shaded regions show $1\sigma$ and $3\sigma$ uncertainties, respectively.
  \label{fig:fdm}
\end{figure}

\subsection{Dark matter fractions}
\label{subsec:fdm}

The formation of the core for the intermediate stellar mass bins, where the stellar mass fractions to the total halo masses are large, can be attributed to the baryonic feedback. To quantitatively investigate this effect, we examine the dark matter fraction within $5r_\mathrm{e}$ defined by
\begin{equation}
  f_{\mathrm{DM}}(<5r_\mathrm{e}) =
\frac{M_{\mathrm{DM,3D}}(<5r_\mathrm{e})}
     {M_{\star,\mathrm{3D}}(<5r_\mathrm{e}) + M_{\mathrm{DM,3D}}(<5r_\mathrm{e})},
\end{equation}
which are shown in Table~\ref{table:fdm} and Figure~\ref{fig:fdm}. Similarly to the results from the previous work based on globular cluster kinematics and orbit-based dynamical models  \citep{2017MNRAS.468.3949A,2024MNRAS.528.5295Y}, we find that the dark matter fraction in the stellar mass bins where cored central density profiles are observed is lower than the values predicted by the EAGLE \citep{2015MNRAS.446..521S,2015MNRAS.450.1937C} and TNG-100 simulations \citep{2018MNRAS.477.1206N,2018MNRAS.480.5113M,2018MNRAS.475..648P,2018MNRAS.475..624N,2018MNRAS.475..676S}. This result suggests that the actual feedback effect may be stronger than predicted by the simulations. 

As partly shown in Figure~\ref{fig:fdm}, dark matter fractions that are lower than predictions of hydrodynamical simulations at centers of early-type galaxies are consistent with several previous dynamical studies \citep[e.g.,][]{2003Sci...301.1696R,2012ApJ...748....2D,2013MNRAS.432.1709C,2017MNRAS.468.3949A,2020MNRAS.491.1690J,2024MNRAS.527..706Z,2024MNRAS.528.5295Y,2024MNRAS.529.4633L} as well as X-ray observations \citep[e.g.,][]{2020ApJ...905...28H}. Interestingly, these previous studies consistently suggest a dip of the dark matter fraction as a function of the stellar mass at around $\sim 10^{11}M_\odot$, which is also seen in our stacked weak lensing result shown in Figure~\ref{fig:fdm}. Moreover, we note that similar results on the dip of the dark matter fraction are obtained from the rotation-curve analysis of late-type galaxies \citep[e.g.,][]{2019MNRAS.489.5483T,2026PASJ...78..745H}. A caveat is that these dynamical studies mostly focus on galaxies at low redshift, $z\sim 0$, while our study constrains the dark matter fraction of early-type galaxies at $z\sim 0.5$. The difference of the redshifts may affect quantitative comparisons.

\subsection{Stellar-to-halo mass relation}
\label{subsec:SHMR}

\begin{figure}[tbp]
  \includegraphics[width=\linewidth]{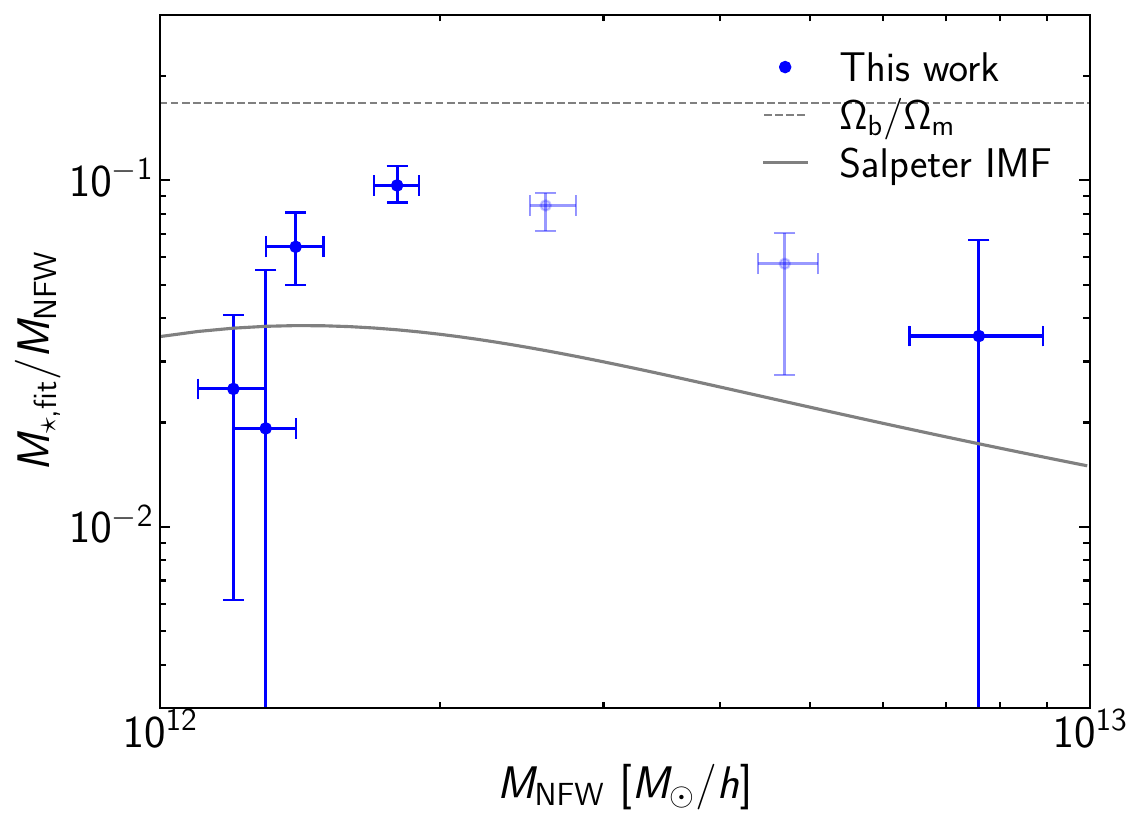}
  \caption{The constraint on the SHMR. The vertical and horizontal axes indicate the stellar mass $M_{\star,\mathrm{fit}}$ obtained from the inner profile fitting normalized by  the halo-mass $M_{\mathrm{NFW}}$ obtained from the outer profile fitting and $M_{\mathrm{NFW}}$, respectively. The blue points represent our results and the solid line shows the case assuming the Salpeter IMF \citep{2019MNRAS.488.3143B,2025OJAp....8E...8A}. A horizontal dotted line shows $\Omega_\mathrm{b}/\Omega_\mathrm{m}$.}
  \label{fig:SHMR}
\end{figure}

In Section~\ref{subsec:inner}, we constrain the stellar mass from the inner profile fitting, and in Section~\ref{subsec:NFW}, we constrain the NFW halo mass by fitting the outer profile. These results allow us to constrain the Stellar-to-halo mass relation (SHMR). In this approach, unlike the conventional approach that estimates stellar masses from spectral energy density fitting, the SHMR is constrained using only weak gravitational lensing. The result is shown in Figure~\ref{fig:SHMR}. Compared to the case assuming the Salpeter IMF \citep{2019MNRAS.488.3143B,2025OJAp....8E...8A}, the stellar mass for a given halo-mass tends to be higher. 
Our result shown in Figure~\ref{fig:min} indicates that the higher normalization of the SHMR in our result is almost entirely driven by the fact that stellar masses measured by stacked weak lensing are a factor of $\sim 2$ higher at $M_{\star,\mathrm{fit}}\gtrsim 10^{11}M_\odot$ than those derived from the spectral energy density fitting assuming the Salpeter IMF. Put another way, the SHMR derived in this paper is quite consistent with the SHMR derived in \citet{2019MNRAS.488.3143B} once the stellar IMF is matched, which assures the validity of our halo mass measurements by stacked weak gravitational lensing.

Therefore our result suggests the possibility of the stellar IMF that is more bottom-heavy than the Salpeter IMF. The bottom-heavy stellar IMF in massive early-type galaxies is qualitatively consistent with previous findings based on spectral absorption lines \citep[e.g.,][]{2010Natur.468..940V}, stellar kinematics \citep[e.g.,][]{2012Natur.484..485C}, and quasar microlensing \citep[e.g.,][]{2014MNRAS.439.2494O,2014ApJ...793...96S}. The stellar IMF derived by strong lensing also tends to prefer the bottom-heavy IMF \citep[e.g.,][]{2010ApJ...709.1195T,2010ApJ...721L.163A,2015ApJ...800...94S,2021MNRAS.503.2380S,2025MNRAS.541....1S,2025A&A...697A..95S}, but with some exceptions \citep[e.g.,][]{2015MNRAS.449.3441S,2019A&A...630A..71S} possibly due to the complex dependence of the stellar IMF on the mass and redshift of galaxies \citep[e.g.,][]{2015MNRAS.449.3441S,2012Natur.484..485C,2024ApJ...973L..32V}. 

Very recently, \citet{2026arXiv260120864C} analyze near-infrared spectra of nine massive, quiescent galaxies at $z\sim 0.7$ with the James Webb Space Telescope to find that their mass-to-light ratios are up to a factor of $\sim 4$ higher than that of the Milky Way, which translates into up to a factor of $\sim 2$ higher stellar masses than those derived assuming the Salpeter IMF. Their results, which study galaxies at redshifts similar to those of our lens sample, are therefore in good agreement with our result.

\section{Conclusion}
\label{sec:concl}

We have investigated stellar mass and central dark matter density profiles of photometric LRGs in seven stellar mass bins in the range $10^{10.3}M_\odot<M_{\star,\,\mathrm{in}}<10^{11.7}M_\odot$ with the stacked weak gravitational lensing analysis of the Subaru HSC-SSP data.
For each stellar mass bin, the inner part of the differential surface mass density profile derived from the weak lensing measurements has been fitted with a two-component model consisting of stellar and dark matter components.
For the dark matter component, we have considered a model that has a central core and follows a power-law in the outer region.

We have found that the core radii of dark matter distributions are consistent with zero within $2\sigma$ for most stellar mass bins, while non-zero core radii have been detected for two stellar mass bins in the range $10^{10.7}M_\odot<M_{\star,\,\mathrm{in}}<10^{11.1}M_\odot$. We have also compared the dark matter profiles constrained from fitting of inner lensing profiles with the extrapolation from fitting of outer lensing profiles to an NFW profile. We have found that they agree within $\sim2\sigma$ for all stellar mass bins except the two stellar mass bins for which non-zero core radii are preferred.

A possible origin of such cores in dark matter distributions for a particular range of stellar masses is the baryonic feedback, as suggested by hydrodynamical simulations. To quantify this feedback effect, we have examined the dark matter fraction within $5r_\mathrm{e}$ to find that, for the two stellar mass bins exhibiting cores, the observed dark matter fractions are lower than those predicted by the EAGLE and TNG-100 simulations.
Our result is broadly consistent with previous measurements of dark matter fractions with stellar dynamics, and suggests that the actual feedback effect may be stronger than that predicted by the simulations.

Furthermore, we provide a new constraint on the SHMR, where both stellar and halo masses are for the first time directly constrained by weak gravitational lensing. Our results indicate relatively high stellar masses for a given halo mass, and prefer the stellar IMF that is more bottom-heavy than the Salpeter IMF.

Our results have demonstrated that stacked weak lensing is a viable tool to study central density profiles of galaxies and already shed new light on central dark matter profiles, the galaxy-halo connection, and the stellar IMF. In addition to applying the methodology developed in this paper to future galaxy survey data, it is important to extend our work by e.g., considering more flexible dark and stellar mass profiles, studying the redshift dependence, investigating in more detail how the results depend on the choice of the fitting range, and exploring the possible effect of the scatter of density profiles among different galaxies, which deserve further exploration in future work. Furthermore, combining weak lensing constraints with those from strong gravitational lensing and stellar dynamics measurements will lead to even tighter constraints on the central dark matter distribution.

\section*{Acknowledgments}

We thank the anonymous referee for useful comments that helped improve the presentation of the paper.
We thank Meng Yang for kindly sharing the numerical data that are used in Figure~\ref{fig:fdm}.
This work was supported by JSPS KAKENHI Grant Numbers JP25H00662, JP25H00672. 

The Hyper Suprime-Cam (HSC) collaboration includes the astronomical communities of Japan and Taiwan, and Princeton University.  The HSC instrumentation and software were developed by the National Astronomical Observatory of Japan (NAOJ), the Kavli Institute for the Physics and Mathematics of the Universe (Kavli IPMU), the University of Tokyo, the High Energy Accelerator Research Organization (KEK), the Academia Sinica Institute for Astronomy and Astrophysics in Taiwan (ASIAA), and Princeton University.  Funding was contributed by the FIRST program from the Japanese Cabinet Office, the Ministry of Education, Culture, Sports, Science and Technology (MEXT), the Japan Society for the Promotion of Science (JSPS), Japan Science and Technology Agency  (JST), the Toray Science  Foundation, NAOJ, Kavli IPMU, KEK, ASIAA, and Princeton University.
 
This paper is based on data collected at the Subaru Telescope and retrieved from the HSC data archive system, which is operated by Subaru Telescope and Astronomy Data Center (ADC) at NAOJ. Data analysis was in part carried out with the cooperation of Center for Computational Astrophysics (CfCA) at NAOJ.  We are honored and grateful for the opportunity of observing the Universe from Maunakea, which has the cultural, historical and natural significance in Hawaii.
 
This paper makes use of software developed for Vera C. Rubin Observatory. We thank the Rubin Observatory for making their code available as free software at http://pipelines.lsst.io/. 
 
The Pan-STARRS1 Surveys (PS1) and the PS1 public science archive have been made possible through contributions by the Institute for Astronomy, the University of Hawaii, the Pan-STARRS Project Office, the Max Planck Society and its participating institutes, the Max Planck Institute for Astronomy, Heidelberg, and the Max Planck Institute for Extraterrestrial Physics, Garching, The Johns Hopkins University, Durham University, the University of Edinburgh, the Queen’s University Belfast, the Harvard-Smithsonian Center for Astrophysics, the Las Cumbres Observatory Global Telescope Network Incorporated, the National Central University of Taiwan, the Space Telescope Science Institute, the National Aeronautics and Space Administration under grant No. NNX08AR22G issued through the Planetary Science Division of the NASA Science Mission Directorate, the National Science Foundation grant No. AST-1238877, the University of Maryland, Eotvos Lorand University (ELTE), the Los Alamos National Laboratory, and the Gordon and Betty Moore Foundation.

\bibliographystyle{aasjournal}

\bibliography{ref}

\begin{appendix}

\section{Tests of systematics in weak lensing measurements}
\label{ap:ap}

\subsection{Background galaxy selection}
\label{ap:P}
\begin{figure}[tbp]
  \includegraphics[width=0.32\linewidth]{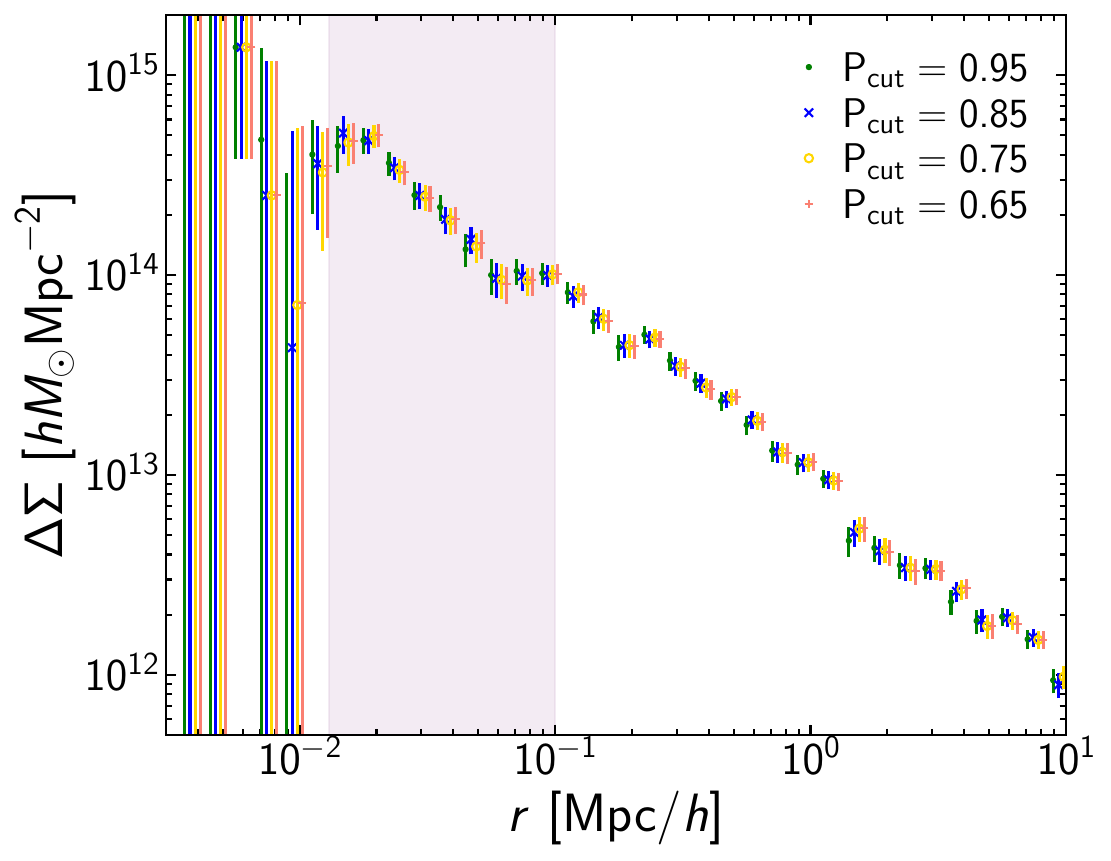}
  \includegraphics[width=0.32\linewidth]{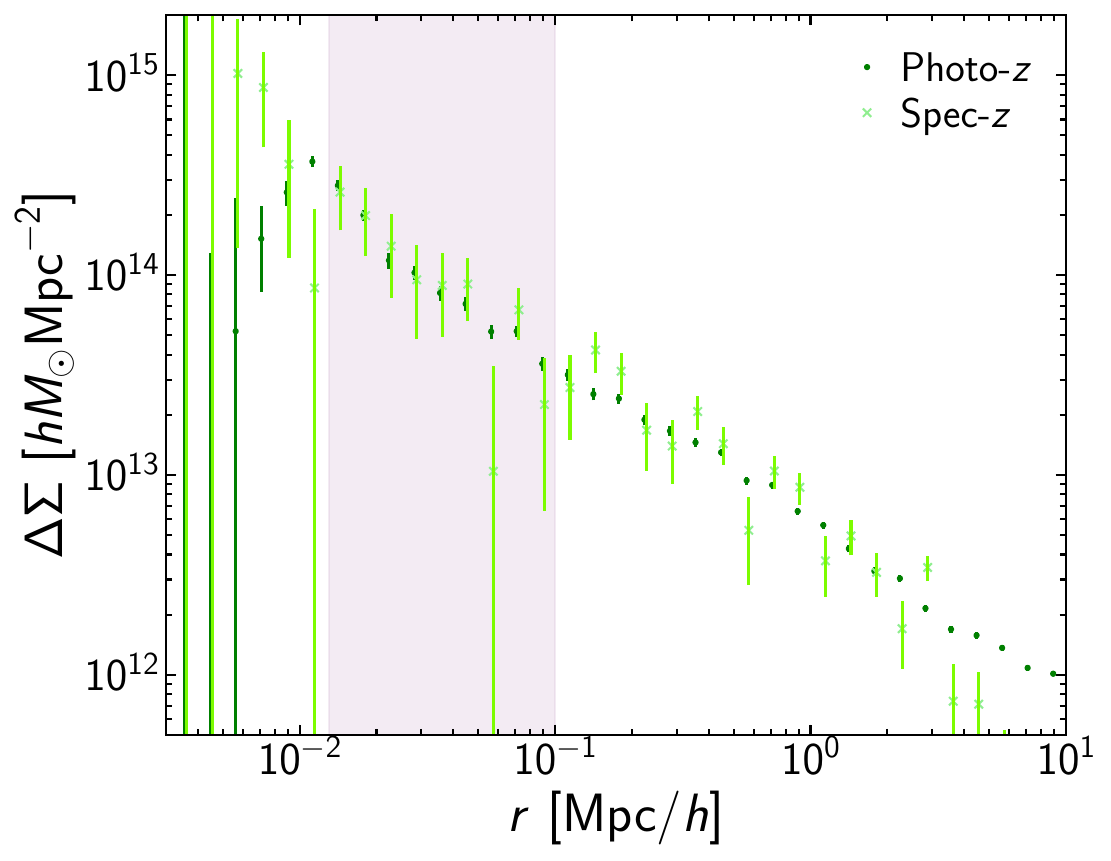}
  \includegraphics[width=0.32\linewidth]{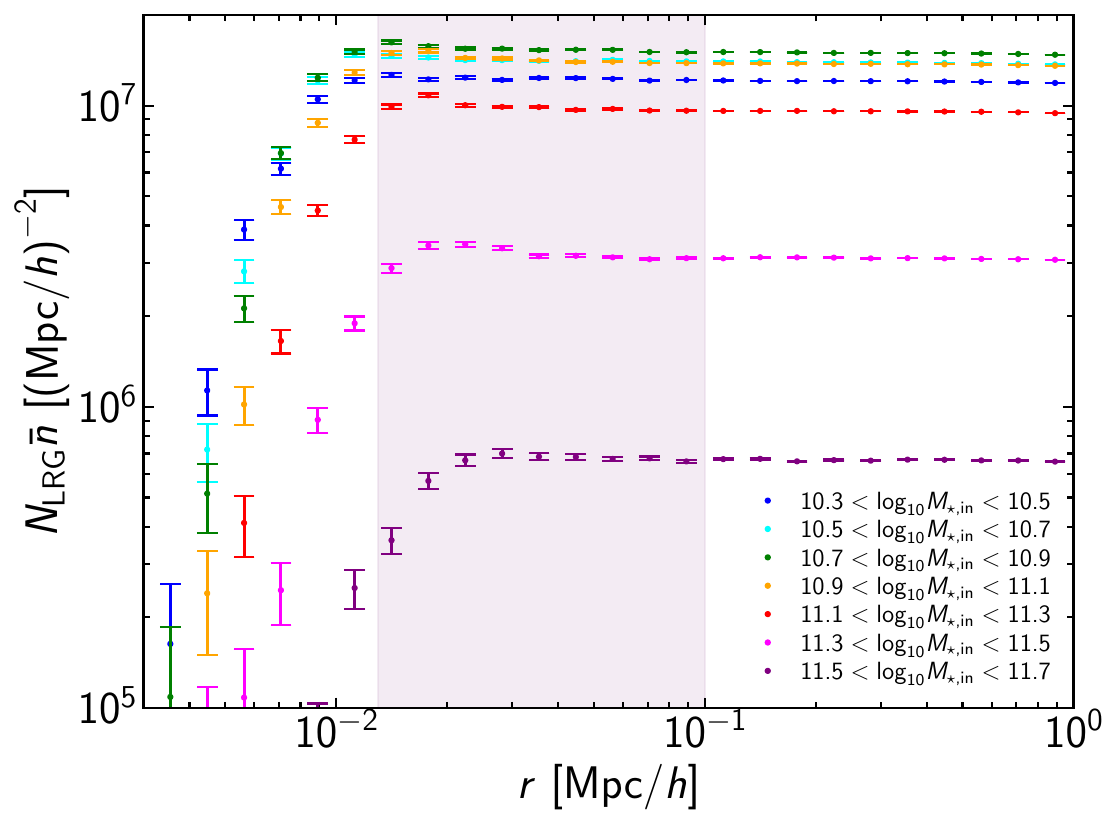}
  \caption{{\it Left:} Dependence of lensing profiles on variations in $P_\mathrm{cut}$ for the stellar mass bin of $10^{11.5}M_\odot<M_{\star,\mathrm{in}}<10^{11.7}M_\odot$. {\it Middle:} Comparison between the results obtained using photometric and spectroscopic redshifts of LRGs for the stellar mass bin of $10^{10.9}M_\odot<M_{\star,\mathrm{in}}<10^{11.1}M_\odot$. In both left and right panels, some points are shifted horizontally for illustrative purpose. {\it Right:} The number density profiles of source galaxies around lensing galaxies for all the seven stellar mass bins considered in this paper. For all the three panels, the shaded region shows the radial range of the inner profile fitting.}
  \label{fig:Pcut}
\end{figure}

We check the validity of our choice of the value of $P_\mathrm{cut}$ in Equation~\eqref{eq:Pcut} by checking how measured lensing profiles are affected by the value of $P_\mathrm{cut}$. The result for the stellar mass bin of $10^{11.5}M_\odot<M_{\star,\mathrm{in}}<10^{11.7}M_\odot$, which corresponds to the highest stellar mass bin and therefore is expected to be most significantly affected by the dilution effect, is shown in the left panel of Figure~\ref{fig:Pcut}. Since the lensing signal becomes unstable in the range $r \le 0.013 h^{-1}\mathrm{Mpc}$, we exclude that from the fitting range. We have confirmed that, for the other stellar mass bins, the results are also stable against variations in $P_\mathrm{cut}$ in the range $r > 0.013h^{-1}\mathrm{Mpc}$.

\subsection{Effect of the photometric redshift uncertainty of LRGs}
\label{ap:spec}

To examine the impact of the photometric redshift uncertainty of photometric LRGs, we compare our results with those using LRGs with spectroscopic redshifts. To do so, we first cross match the photometric LRGs with a compilation of spectroscopic galaxies constructed in \citet{2026PASJ..tmp....2O}, which is used to calibrate the stellar population synthesis model in the CAMIRA algorithm. The majority of the spectroscopic galaxies at $0.4<z<0.6$ are those from the SDSS \citep{2019ApJS..240...23A} and DESI \citep{2025arXiv250314745D}. We then rerun the algorithm to select photometric LRGs with redshifts of cross-matched galaxies fixed to the spectroscopic redshifts to re-compute their stellar masses. The number of these spectroscopic LRGs is roughly a factor of $10$ smaller than that of the photometric LRGs. The result for the stellar mass bin $10^{10.9}M_\odot<M_{\star,\mathrm{in}}<10^{11.1}M_\odot$ is shown in the middle panel of Figure~\ref{fig:Pcut}.
Defining the chi-square as
\begin{equation}
  \chi^2 = \sum_{i=1}^{N} \frac{(\Delta\Sigma_{\mathrm{spec},i} - \Delta\Sigma_{\mathrm{photo},i})^2}{\sigma_{\mathrm{spec},i}^2 + \sigma_{\mathrm{photo},i}^2},
  \label{eq:chispec}
\end{equation}
we obtain $\chi^2 \approx 4.6$ within the fitting range for the inner density profile ($N=9$ bins) for this stellar mass bin. Here, $\Delta\Sigma_{\mathrm{photo}}$, $\sigma_{\mathrm{photo}}$ and $\Delta\Sigma_{\mathrm{spec}}$, $\sigma_{\mathrm{spec}}$ are the differential surface mass densities and their errors derived from the photometric and spectroscopic LRGs, respectively. Since the value of $\chi^2$ lies within the $95\%$ confidence interval ($2.7\lesssim\chi^2\lesssim19.0$), the difference of lensing profiles between the photometric and spectroscopic LRGs is not statistically significant.
Therefore, we conclude that the effect of the photometric redshift uncertainty of LRGs is negligible. We have computed $\chi^2$ for other stellar mass bins as well, and for the stellar mass bin $10^{11.3}M_\odot<M_{\star,\mathrm{in}}<10^{11.5}M_\odot$, we find that $\chi^2 \approx 22.8$ that falls outside the $95\%$ confidence interval. Therefore, the data in this bin are less reliable and should be interpreted with caution. We find that the values of $\chi^2$ for the other mass bins are well within $2.7\lesssim\chi^2\lesssim19.0$, indicating that the differences of lensing profiles between the photometric and spectroscopic LRGs are also negligible for those stellar mass bins.

\subsection{Cross shear}
\label{ap:cross}
The cross shear profile should be consistent with zero in the absence of any systematic errors, and hence is used to check systematic errors in tangential shear profile measurements. We define the chi-square as
\begin{equation}
  \chi^2 = \sum_{i=1}^{N} \frac{\Delta\Sigma_{\times,i}^2}{\sigma_{\times,i}^2},
  \label{eq:chicross}
\end{equation}
where $\Delta\Sigma_{\times}$ and $\sigma_{\times}$ are the differential surface mass density and its error derived from the cross shear, respectively. For the fitting range for the inner density profile ($N=9$ bins), we obtain $\chi^2$ ranging from $2.2$ to $14.55$ across the stellar mass bins, which mostly lie within the $95\%$ confidence interval ($2.7\lesssim\chi^2\lesssim19.0$). Therefore this cross shear test suggests no sign of systematics for the innermost lensing profiles that are our main interest in this study. We also check the $\chi^2$ values for all the fitting range of the inner and outer density profiles ($N=24$ bins). We obtain $\chi^2$ ranging from $14.9$ to $37.4$ across the stellar mass bins, which again lie within the $95\%$ confidence interval 
($12.4\lesssim\chi^2\lesssim39.4$).

\subsection{Number counts of source galaxies around lensing galaxies}

The number density profiles of source galaxies around lensing galaxies are affected by several effects such as the dilution effect due to galaxies physically associated with lensing galaxies, the magnification bias, and the obscuration effect by lensing galaxies, and serve as a check of these effects. Among these, the dilution effect increase the number density profiles toward the small radii and leads to the underestimation of weak lensing signals, while the impacts of the magnification bias and the obscuration effect on lensing signals are complicated and depend on the situation. The right panel of Figure~\ref{fig:Pcut} shows the number density profiles for all the seven stellar mass bins. We find that the profiles are flat for low stellar mass bins, which indicate that there is no obvious sign of these systematic effects affecting the observed lensing signals. On the other hand, it is seen that the innermost radial bins appear to be partly affected the obscuration effect for the two highest stellar mass bins, which is judged based on the rapid decreases of the number density profiles. The larger obscuration effect for the higher stellar mass bins is explained by the brighter lensing galaxies. While the best-fitting results shown in Figure~\ref{fig:inner fit} suggest that any bias in weak lensing measurements in the innermost radial bins for the two highest stellar mass bins, if any, mainly affect their stellar mass measurements, large measurement errors on stellar masses for these stellar mass bins indicate that its impact on the result is likely to be insignificant. We comment that the cross shear test conducted above shows no sign of significant systematic errors in weak lensing shape measurements even for these highest stellar mass bins. We also note that, for these highest stellar mass bins, there is a slight increase of the number density profiles with decreasing radii. However the level of the increase is $\lesssim\,10\%$ at most, which affect e.g., the stellar mass measurement also at the $\lesssim 10\%$ level for the highest stellar mass bins, which is not large enough to change our conclusion.

We also check the obscuration effect using surface brightness distribution measurements of quiescent galaxy in \citet{2025ApJ...989..107W}. For quiescent galaxies at $z\sim 0.5$ and stellar masses of $M_\star\geq 10^{11}M_\odot$, they measure the average HSC $i$-band surface brightness values of $\sim 26~\mathrm{mag}/\mathrm{arcsec}^2$ at $r=4r_{\mathrm{e}}$, which corresponds to the inner boundary radius of the inner profile fitting for the highest stellar mass bin (see Table~\ref{table:in}), and $\sim 29~\mathrm{mag}/\mathrm{arcsec}^2$ at $r=8r_{\mathrm{e}}$, which corresponds to the typical inner boundary radius of the inner profile fitting (see Table~\ref{table:in}). Even if we assume a conservative size of source galaxies of $\sim 1''$, the contamination of the lens light is at most $\sim 26$~mag, which is much fainter than magnitudes of source galaxies, $i\leq 24.5$~mag \citep{2022PASJ...74..421L}. This indicates that the obscuration effect by foreground lensing galaxies appears to be typically insignificant, which is consistent with the number count analysis presented in Figure~\ref{fig:Pcut}.

We caution that the more quantitative assessment of these effects requires more careful analysis. For instance, the assessment of the obscuration effect ultimately requires e.g., intensive injection simulations of fake galaxies and run the HSC-SSP image analysis pipeline to see how lensing and source galaxies are deblended and how shape measurements of source galaxies are affected. The huge number of lensing galaxies for our work, however, makes such injection simulations highly challenging. We leave such exploration to future work.

\end{appendix}

\end{document}